\documentclass[twocolumn,prd,nofootinbib,superscriptaddress,longbibliography]{revtex4-1} %
\newcommand\ForInternalReference[1]{}
\newcommand\SkipForEarlyCirculation[1]{}

\newcommand\SkipPP[1]{}

\usepackage[colorlinks=true,citecolor=blue,urlcolor=blue]{hyperref}
\usepackage{verbatim}
\usepackage{graphicx}
\usepackage{dcolumn}
\usepackage{bm}
\usepackage{xcolor}
\usepackage{url}
\usepackage{float}
\usepackage{multirow}
\usepackage{amssymb}
\usepackage{adjustbox}
\usepackage[normalem]{ulem}
\usepackage{bm}
\usepackage{placeins}
\usepackage{afterpage}
\usepackage{amsmath,amssymb}
\usepackage{acronym}
\DeclareUnicodeCharacter{2299}{\ensuremath{\odot}}
\usepackage{booktabs}      
\usepackage{tabularx}      
\usepackage{array}         

\newcommand\optional[1]{}
\acrodef{NR}[NR]{Numerical Relativity}

\definecolor{amber}{rgb}{1.0, 0.75, 0.0}
\definecolor{orange}{rgb}{1.0, 0.5, 0.0}
\definecolor{amaranth}{rgb}{0.9, 0.17, 0.31}

\graphicspath{{./figures/}}

\def\ltsima{$\; \buildrel < \over \sim \;$}
\def\simlt{\lower.5ex\hbox{\ltsima}}
\def\gtsima{$\; \buildrel > \over \sim \;$}
\def\simgt{\lower.5ex\hbox{\gtsima}}

\def\eos#1{equation of state#1 (EOS#1)\gdef\eos{EOS}}
\def\QNM#1{quasi-normal mode#1 (QNM#1)\gdef\QNM{QNM}}
\def\ns#1{neutron star#1 (NS#1)\gdef\ns{NS}}
\def\gw#1{gravitational wave#1 (GW#1)\gdef\gw{GW}}
\def\bh#1{black hole#1 (BH#1)\gdef\bh{BH}}
\def\bbh#1{binary black hole#1  (BBH#1)\gdef\bbh{BBH}}
\def\bns#1{binary neutron star#1 (BNS#1)\gdef\bns{BHS}}
\def\bhns#1{black hole - neutron star#1 (BHNS#1)\gdef\bhns{BHNS}}
\def\nsbh#1{neutron star - black hole#1 (NSBH#1)\gdef\nsbh{NSBH}}
\def\nr#1{numerical relativity#1 (NR#1)\gdef\nr{NR}}

\newcommand{\Austin}{\affiliation{Department of Physics, University of Texas at Austin, Austin, TX 78712, USA}}

\newcommand{\enineteen}{e_\mathrm{19}}

\begin{document}
\title{Impact of eccentricity and higher-modes on neutron star-black hole parameter estimation}
\author{Snehal Tibrewal} \email{snehaltibrewal@gmail.com}
\Austin
\author{Aasim Jan} \Austin
\author{Aaron Zimmerman} \Austin
\author{Hsin-Yu Chen} \Austin

\date{\today}

\begin{abstract}
Detections of gravitational waves from neutron star–black hole systems provide avenues for studying extreme matter, constraining binary formation channels, and testing the nature of compact objects in strong gravity. 
Eccentric signatures in the signal further enhance this potential by improving parameter estimation and offering clues about binary formation. 
Because eccentricity is primarily imprinted during the inspiral phase, it is often weakly constrained or missed entirely in binary black hole observations; in contrast, neutron star-black hole systems produce longer in-band signals, enabling more precise measurements of eccentricity and leaving a distinct imprint on parameter inference. 
In this work, we present a systematic parameter-estimation study exploring the impact of eccentricity on inference using injections simulated with the state-of-the-art eccentric waveform model {\tt SEOBNRv5EHM}. 
We find that for systems like GW200105\_162426, the measurement precision of eccentricity and correlated parameters improves as eccentricity increases, yielding tighter constraints at larger eccentricities. For the highest eccentricity considered in this study, $e=0.25$, we recover eccentricity with $1\sigma$ uncertainty as low as $4\times10^{-4}$. 
In addition, the constraints on effective spin $\chi_\mathrm{eff}$ and mass ratio $q$ improve relative to the quasi-circular case by factors of $\sim13$ and $\sim20$, respectively. 
On the other hand, we find no significant improvement in extrinsic parameters such as luminosity distance and sky localization, suggesting that for systems like GW200105\_162426, the additional information provided by eccentricity in this sector is either negligible or degenerate with the information provided by higher-order modes.
\end{abstract}

\maketitle

\section{Introduction}

Merging neutron star-black hole (NSBH) binaries are strong sources of gravitational waves (GWs)~\cite{LIGOScientific:2021qlt}, and also potential multi-messenger sources~\cite{Foucart:2020ats}. 
They provide probes of the physics of compact objects, strong-field gravity~\cite{Foucart:2018rjc}, and the origins of heavy elements when tidally disrupted~\cite{ Metzger:2019zeh}. 
So far, the LIGO-Virgo-KAGRA~\cite{LIGOScientific:2014pky,VIRGO:2014yos, KAGRA:2018plz} network has confidently identified four signals consistent with NSBH-like systems~\cite{LIGOScientific:2021qlt, LIGOScientific:2024elc, LIGOScientific:2025slb}. These can be considered the four principal NSBH discoveries although the broader catalog contains additional lower significance, classification-dependent candidates. 
One particular event, GW200105\_162426~\cite{LIGOScientific:2021qlt}, hereafter GW200105, was detected in 2020 during the third observing run (O3) by the LIGO Livingston and Virgo detectors. The public strain data for this event are available through the GWOSC event page \href{https://gwosc.org/eventapi/html/GWTC-3-marginal/GW200105_162426/v2/}{here}, as part of the O3 open-data release~\cite{KAGRA:2023pio}. The signal had a network signal-to-noise ratio (SNR) of approximately 13.4 and a false alarm rate (FAR) of about 1 per 2.8 years in the original quasi-circular searches. The FAR was later updated to be less than 1 per 1000 years with a targeted eccentric search by Phukon et al.~\cite{Phukon:2026cmtb}. 
Its source is consistent with an NSBH binary with the source-frame component masses of about $8.9^{+1.2}_{-1.5} M_\odot$ and $1.9^{+0.3}_{-0.2} M_\odot$, located at a luminosity distance of roughly 280 Mpc. 
This event has received significant renewed interest due to several recent analyses reporting evidence for measurable orbital eccentricity in the binary~\cite{morras2025orbitaleccentricityneutronstar, planas2025eccentricinspiralmergerringdownanalysisneutron, Jan:2026zjmc, kacanja2025eccentricitysignaturesligovirgokagrasbns, Phukon:2026cmtb, Pompili:2026yxq}. 
This is an intriguing development as measurement of orbital eccentricity provides valuable information about the origins and formation history of binaries of compact objects such as black holes and neutron stars, which continue to be a major scientific mystery in astronomy. Previously, there have been other compact binary candidates, specifically binary black holes, that have been classified as eccentric~\cite{Romero-Shaw:2022xko, Iglesias:2022xfc, Gupte:2024jfe}

Isolated binary evolution is commonly considered a leading formation channel for NSBH systems~\cite{Belczynski_2002,Broekgaarden_2021,Mandel_2021}. 
This channel is generally expected to produce systems with low residual eccentricity by the time they enter the LVK band~\cite{Peters_1964,Broekgaarden_2021}. 
The reported evidence of eccentricity in the source of GW200105 sparks the possibility that the binary formed through a different path~\cite{Romero-Shaw:2025otx}. 
Eccentric binaries could be formed through dynamical channels. 
For example, in hierarchical triple systems, interactions with a distant third companion can transfer angular momentum to the inner binary and instigate orbital eccentricity \cite{Silsbee_2017,Antonini_2017,Fragione_2019_1,1910AN....183..345V,Lidov:1962wjn,1962AJ.....67..591K}. 
Eccentricity can also arise when binaries are assembled through close encounters in dense stellar environments, where repeated interactions and exchanges can result in the formation of a tight binary with large eccentricity, which does not fully circularize before merger~\cite{Freire_2004,Fragione_2019,Ye_2019,Rastello_2020,Trani_2022}.
Despite their different astrophysical origins, these dynamical channels may produce a common eccentricity distribution in the high-eccentricity regime~\cite{Rozner:2026jtj}.
The recent identification of eccentricity in a NSBH suggests that such dynamical formation channels may produce NSBHs, and motivates continued searches for eccentric signatures in even relatively low-mass GW sources~\cite{Stegmann:2025clo,Romero-Shaw:2025otx}.

In addition to carrying direct information about the dynamical history of the system, eccentricity introduces additional orbital harmonics and modulations into the waveform, enriching its time-frequency structure.
These structural changes can alter correlations between inferred parameters and provide additional information for parameter recovery. 
Previous work has shown that neglecting eccentric orbital dynamics in the recovery waveform can lead to significant systematic biases in compact-binary inference, e.g.~\cite{Wu:2020zwr,Huez:2025npe,RoyChowdhury:2026xgb,Yang:2026mam}. 
Meanwhile when eccentricity is present and incorporated in modeling, it can aid in the inference of source parameters.
Early work done for the space-based detector LISA~\cite{Babak:2021mhe} showed that the additional harmonic structure induced by eccentricity can reduce the uncertainties in intrinsic parameter measurements and break parameter degeneracies~\cite{Mikoczi:2012qy}. 
More recently, studies in the context of ground-based LVK detectors have shown similar effects.
Gondán et al.~\cite{Gondan:2017hbp,Gondan:2018khr} showed through Fisher-matrix studies that the additional harmonic content of eccentric compact-binary signals can improve detectability and parameter-estimation precision in parts of the parameter space. In particular, eccentric structure can weaken parameter degeneracies, improving measurements of intrinsic quantities such as the chirp mass and, for favorable systems, sky localization.
Favata et al.~\cite{Favata:2021vhw} showed that small eccentricities are better constrained in long-inspiral systems, including BNSs and NSBHs, and can correlate with intrinsic parameters through a chirp-mass--eccentricity degeneracy. 
These studies motivate eccentricity as a source of additional information for intrinsic-parameter recovery. 
More specifically for GW200105-like NSBHs, Morras et al.~\cite{morras2025orbitaleccentricityneutronstar} found that an eccentric-precessing analysis of GW200105 produced narrower posteriors for the component masses and mass ratio compared to non-eccentric analyses.
A related set of studies has focused on source localization and early-warning observations. Eccentric binaries contain additional orbital harmonics that can enter the detector band earlier than the dominant quadrupolar harmonic, potentially increasing the available warning time and improving sky localization for favorable source configurations~\cite{Yang:2022tig,Ma:2017bux,Pan:2019anf,Yang:2023zxk,Yang:2024vfy, Sinha:2025vmc}. 

Much of the prior work focuses on Fisher-matrix and other semi-analytic approaches, with some supplemented by Bayesian inference using Monte Carlo sampling. These analyses used the eccentric waveform models available at the time, many of which described only the inspiral, retained only the dominant mode, neglected spin precession, or adopted restricted treatments of the orbital phase and mean anomaly. More recent studies have used complete inspiral--merger--ringdown eccentric waveform models to analyze GW200105 and other NSBH events~\cite{planas2025eccentricinspiralmergerringdownanalysisneutron,kacanja2025eccentricitysignaturesligovirgokagrasbns,Jan:2026zjmc, Pompili:2026yxq}. Meanwhile, existing NSBH-specific work has mainly focused on investigating the presence of eccentricity in GW200105 itself, along with the robustness and astrophysical interpretation of the inferred eccentricity~\cite{morras2025orbitaleccentricityneutronstar,planas2025eccentricinspiralmergerringdownanalysisneutron,kacanja2025eccentricitysignaturesligovirgokagrasbns,Jan:2026zjmc,Phukon:2026cmtb,Tiwari:2025fua,Clarke:2026cuw}. Our work investigates the broader impact of eccentricity on the recovery of intrinsic and extrinsic parameters in controlled GW200105-like NSBH systems, which has not yet been systematically quantified.

This paper is organized as follows. 
We describe the technical details regarding the injections and parameter estimation in Sec.~\ref{sec:Methods}. 
We then present the results from our work in context of intrinsic parameters in Sec.~\ref{sec:Intrinsic} and extrinsic parameters in Sec.~\ref{sec:Extrinsic}. 
Our findings from the follow-up investigations surrounding inclination and higher-order modes are detailed in Sec.~\ref{sec:InclinationDependence} and Sec.~\ref{sec:HM}, respectively. 
Finally, in Sec.~\ref{sec:Discussion}, we discuss the implications and robustness of our results and outline directions for future work.

\section{Model, Injections and Parameter Estimation}
\label{sec:Methods}

\begin{table}[tb]
\centering
\renewcommand{\arraystretch}{1.25}
\label{tab:injection_parameters}
\begin{tabular}{|l|l|}
\toprule
\hline
\textbf{Parameter} & \textbf{Value} \\
\midrule
\hline
Primary mass ($m_\mathrm{1,det}$) & $8.74 \, M_\odot$ \\
Secondary mass ($m_\mathrm{2,det}$) & $2.16 \, M_\odot$ \\
Primary aligned spin ($\chi_{1,z}$) & $-0.07$ \\
Secondary aligned spin ($\chi_{2,z}$) & $-0.03$ \\
Right ascension ($\alpha$) & $1.70\,\mathrm{rad}$ \\
Declination ($\delta$) & $-0.13\,\mathrm{rad}$ \\
Polarization angle ($\psi$) & $1.19\,\mathrm{rad}$ \\
Coalescence phase ($\phi_\mathrm{c}$) & $3.18\,\mathrm{rad}$ \\
Coalescence time ($t_\mathrm{c}$) & $1000000000.0$ \\
Relativistic anomaly ($\zeta$) & $0\, \mathrm{rad}$ \\
SNR ($\rho$) & 20.0 \\
\bottomrule
\hline
\end{tabular}
\caption{Fixed parameters across all injections. 
The relativistic anomaly $\zeta$ is defined at the reference frequency, here $f_{\text{ref}} = 19$ Hz.}
\label{table:fixed_param}
\end{table}

For this analysis, we generate theoretical signals using a state-of-the-art waveform model---{\tt SEOBNRv5EHM}~\cite{gamboa2024accuratewaveformseccentricalignedspin}, which includes both the eccentric, aligned-spin dynamics and higher multipolar emission. 
We use higher order modes up to $l_{\text{max}}=4$, which includes these modes offered by the model: $\{(2,\pm2), (2,\pm1), (3,\pm3), (3,\pm2), (4,\pm4), (4,\pm3)\}$. Some of the higher modes used here are redundant as we did find that $\sim98.5\%$ of the signal comes from the $\ell_\mathrm{max}=2$ modes, hence making the higher-mode contribution less significant for this system. 
Consistent with the inferred spins of GW200105 in~\cite{morras2025orbitaleccentricityneutronstar,zjmc-117s}, we assume our systems are non-precessing. While we do not include tidal effects in the model, for GW200105-like systems with $M_{\text{tot}}>10 \, M_\odot$ and mild spins, such effects are expected to be negligible~\cite{LIGOScientific:2021qlt, Huang:2020pba}, making this model suitable for our analysis. 
Our choice for not including tidal effects is further supported by work done by Jan et al.~\cite{zjmc-117s} that reports a non-disruptive merger in numerical relativity simulations of GW200105-like systems. 
We generate the waveforms at the same starting and reference frequency, $f_{\text{min}} =f_{\text{ref}} = 19$ Hz. 
This is the frequency at which all time/frequency varying parameters are defined. 
The lower sensitivity threshold for the detectors in the network is set to $f_\mathrm{low}=20$ Hz. 
Starting the waveforms closer to this threshold implies that some of the previously mentioned higher modes begin in band. 
Ideally $f_{\text{min}}$ would be set such that all higher modes are fully resolved in the likelihood, but this comes at the cost of increased computational expense in generating the time-domain waveform starting from a lower frequency.
We choose to limit the time-domain waveforms to fit within a 32 second duration in order to control the computational expense of our analyses.
Further, since we use the same waveform settings for both injection and PE recovery, we expect the effect of the missing higher mode content to be small.

We anchor our injections to the parameter space surrounding GW200105 and explore the effects of eccentricity on the inference of intrinsic and extrinsic parameters for similar systems. To choose the parameters of our injections, we start with the highest likelihood point from the IMRPhenomNSBH (LowSpin) LVK analysis of GW200105~\cite{GW200105_data, KAGRA:2021vkt}. 
We then create a suite of injections by varying the eccentricity $e \in \{0, 0.1, 0.25\}$. 
Eccentricity and relativistic anomaly are frequency dependent parameters and are reported at the reference frequency ($\enineteen$ and $\zeta_{19}$ hereafter). Values for all other parameters, including the relativistic anomaly $\zeta$, are fixed across all injections (see Table~\ref{table:fixed_param} for the complete list). 
Our fiducial analysis assumes a face-on configuration, with inclination $\iota=0$. For each value of $e_{19}$, we set the luminosity distance such that the optimal network SNR is $\rho_{\rm opt}=20.0$. In Sec.~\ref{sec:Followup}, we explore the dependence on inclination by varying $\iota \in \{0, \pi/6, \pi/3\}$ and adjusting the luminosity distance in each case to keep $\rho_{\rm opt}=20.0$. This fixed-SNR setup allows us to isolate changes in parameter recovery due to eccentricity and inclination from changes caused by signal strength. We also perform injection-recovery runs using only the $\ell_\mathrm{max}=2$ modes, in order to assess the impact of higher modes on parameter measurements. 

We inject these signals into zero-noise detector data, and carry out full Bayesian parameter estimation (PE) using the same model for injection and recovery. PE is conducted using {\tt RIFT} \cite{PhysRevD.92.023002,lange2018rapidaccurateparameterinference,wofford2023expandingriftimprovingperformance,wagner2025narrowingriftfocusedsimulationbasedinference}. {\tt RIFT} is a rapid parameter estimation pipeline that accelerates inference through iterative likelihood interpolation and marginalization over selected extrinsic parameters. Low mass binaries like NSBHs enter the LVK sensitivity band earlier than the more massive BBH systems, acummulating more cycles in-band before merger. As a result, their analysis can be computationally expensive, making {\tt RIFT} well suited for this study.  
For our analysis we assume a detector network consisting of LLO, LHO and Virgo with Design Sensitivity PSDs~\cite{LIGO-T2200043}
and carry out a coherent analysis in all three detectors.
We utilize standard agnostic priors for aligned-spin binaries, as described in Appx.~\ref{sec:Priors}.

\section{Effect of eccentricity on GW parameters}
Here we present the results of our injection and recovery analysis, focusing first on the parameter inference for our fiducial suite of face-on binary signals as we vary eccentricity.

\subsection{Intrinsic parameter measurements}
\label{sec:Intrinsic}

\begin{figure*}[t]
\includegraphics[width=0.325\textwidth]{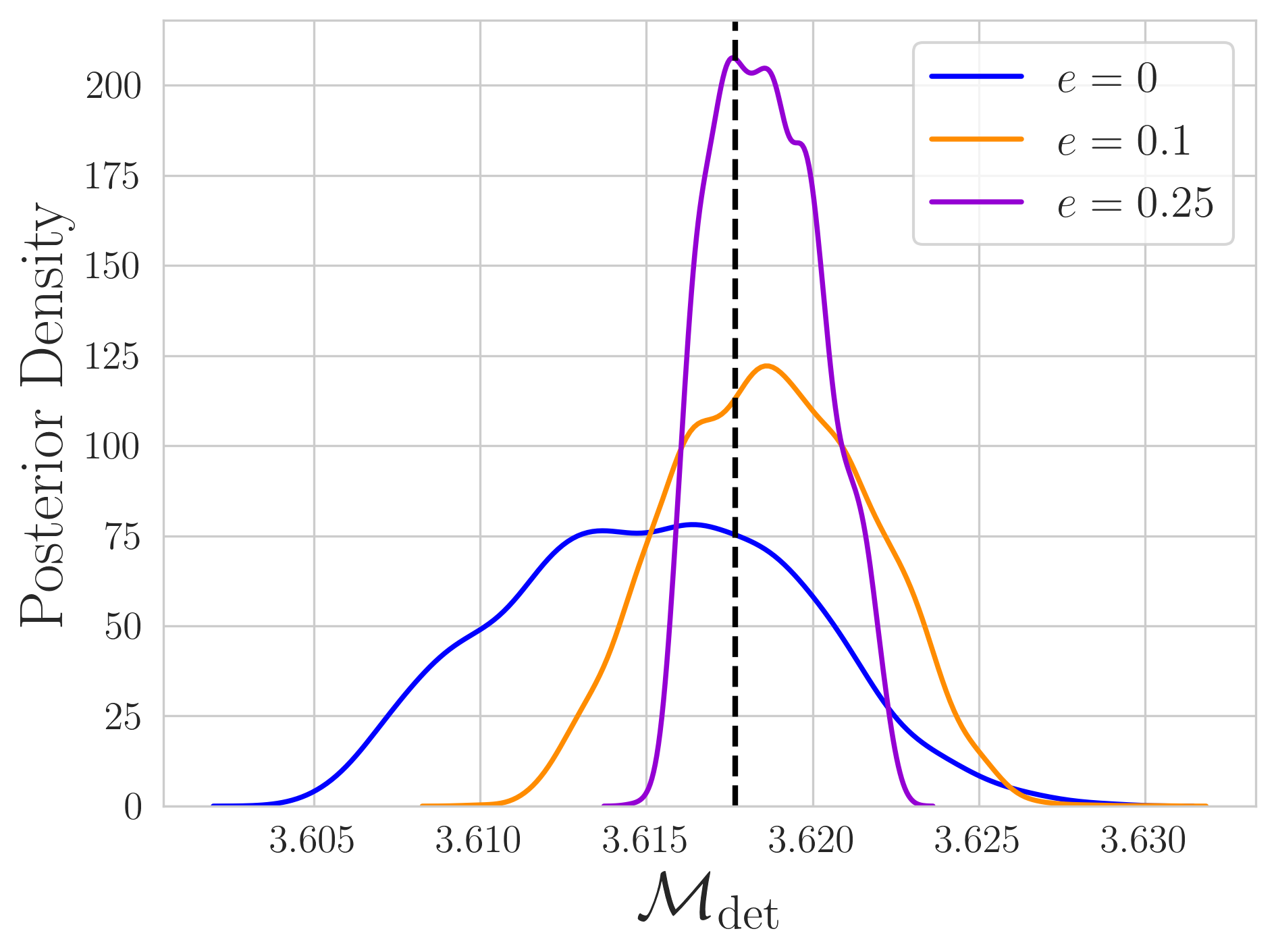}
\includegraphics[width=0.325\textwidth]{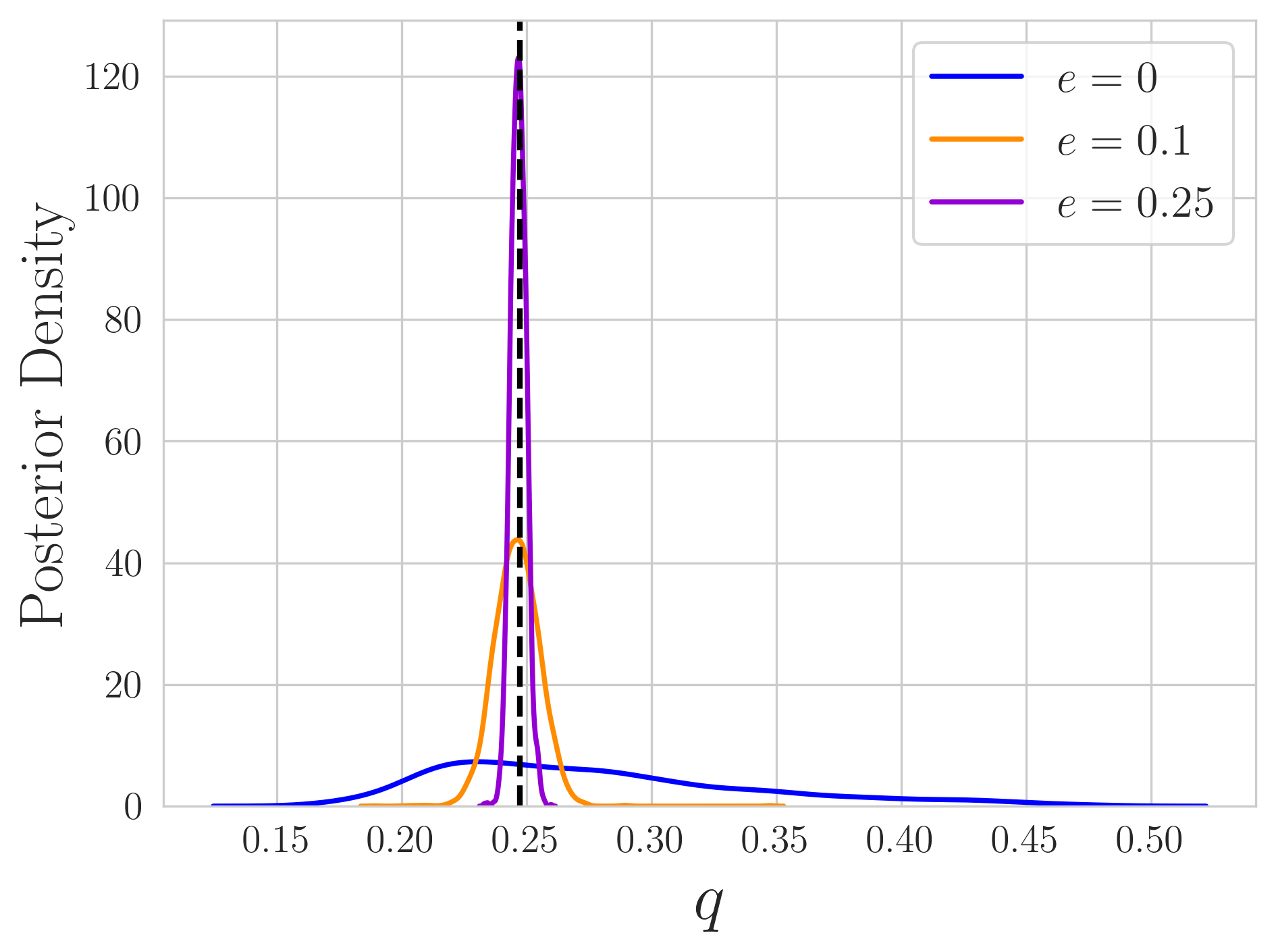}
\includegraphics[width=0.325\textwidth]{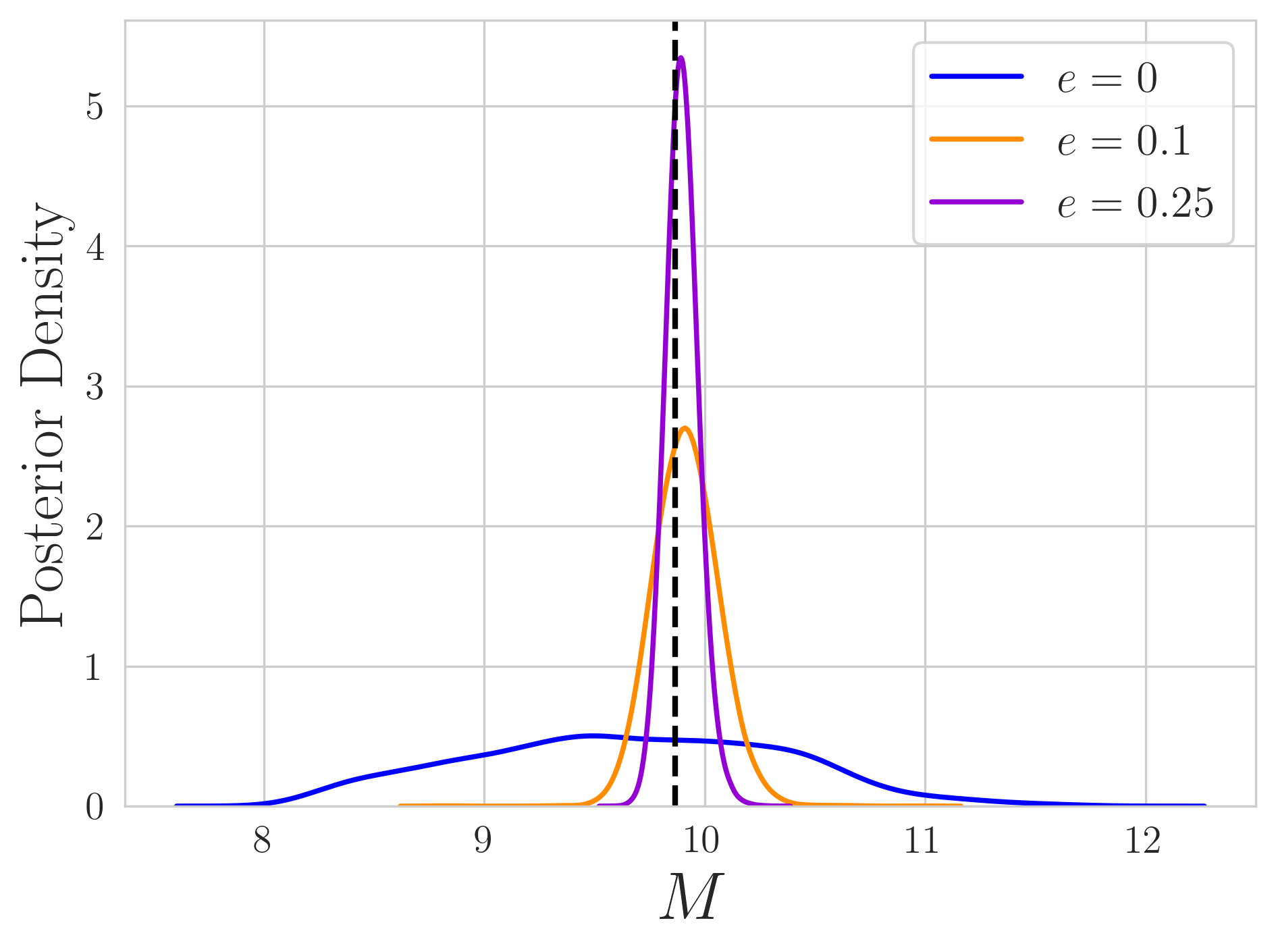}
\includegraphics[width=0.325\textwidth]{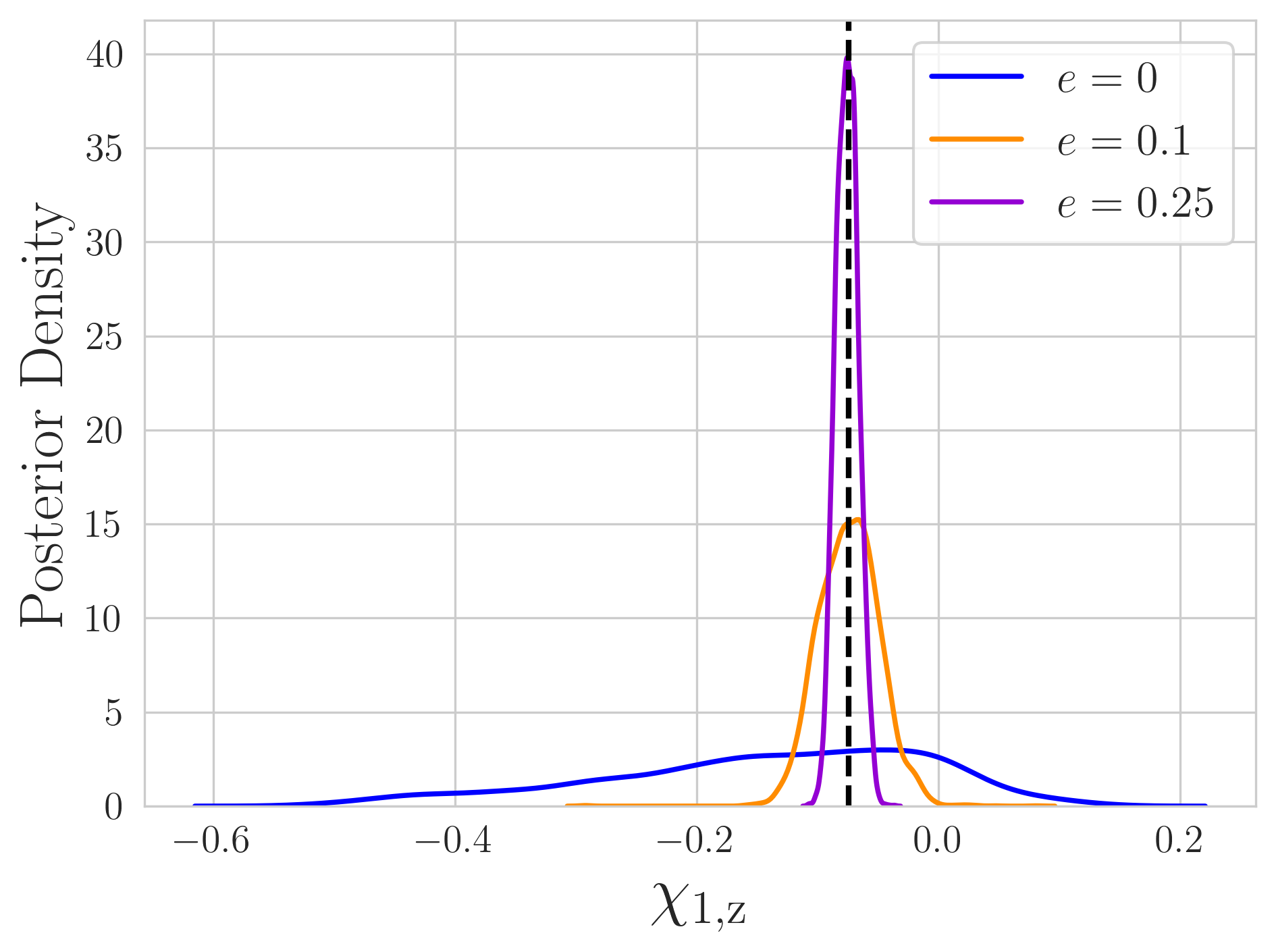}
\includegraphics[width=0.325\textwidth]{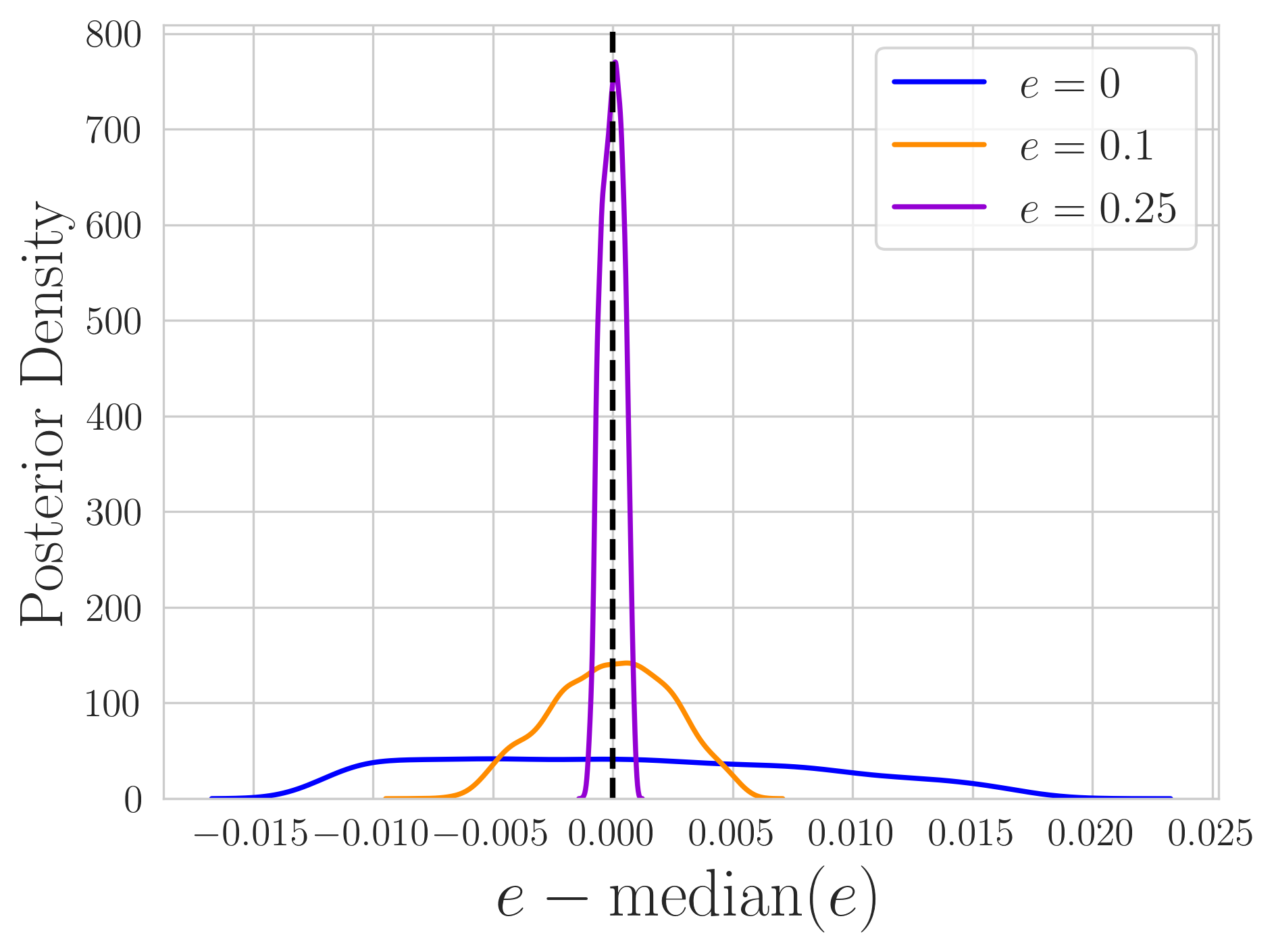}
\includegraphics[width=0.325\textwidth]{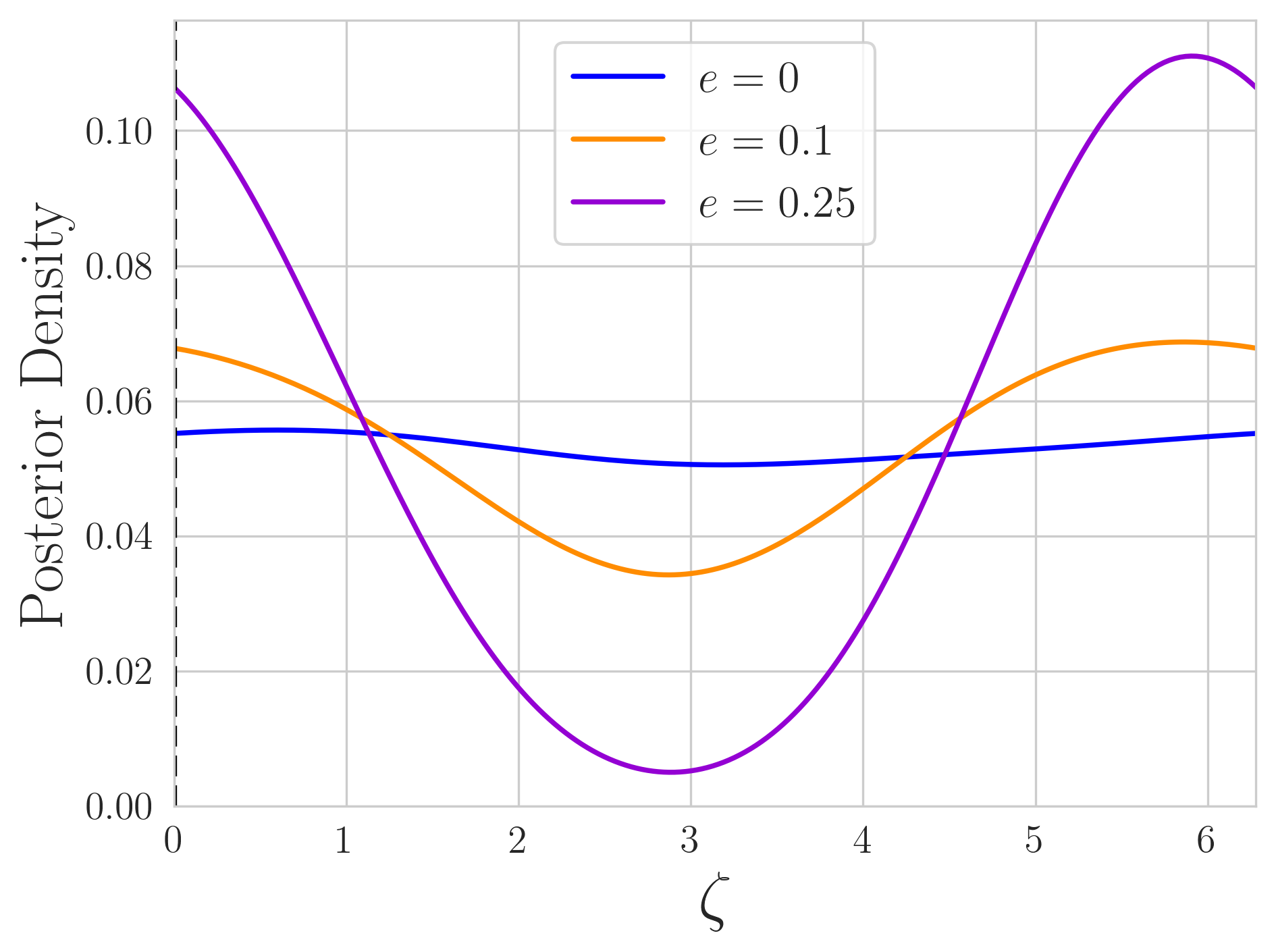}
\caption{1D posteriors for intrinsic parameters compared across all 3 eccentricities: $\enineteen=0$ in blue, $\enineteen=0.1$ in orange and $\enineteen=0.25$ in purple. The true injected value for each parameter is plotted as a black dashed line. The generic trend observed here is the improvement in precision of the recovered intrinsic parameters as the value of injected eccentricity increases.}
\label{fig:1D_intrinsic}
\end{figure*}

\begin{table}[tb]
\centering
\renewcommand{\arraystretch}{1.25}
\begin{tabular}{l | c c}
\hline
\hline
Parameter 
& $\sigma_{e=0}/\sigma_{e=0.1}$ 
& $\sigma_{e=0}/\sigma_{e=0.25}$ \\
\hline
\hline
Eccentricity $(e)$                 & 3.1 & 17.8 \\
Primary mass $(m_{1,\mathrm{det}})$               & 5.3 & 15.7 \\
Secondary mass $(m_{2, \mathrm{det}})$               & 6.5 & 18.9 \\
Total mass $(M_\mathrm{det})$        & 5.1 & 15.0 \\
Chirp mass $(\mathcal{M}_\mathrm{det})$        & 1.5 & 2.7 \\
Source-frame primary mass $(m_{1})$      & 5.2 & 12.0 \\
Source-frame secondary mass $(m_{2})$      & 6.0 & 12.6 \\
Source-frame total mass $(M)$        & 4.7 & 9.2 \\
Mass ratio $(q)$                 & 6.9 & 20.3 \\
Symmetric mass ratio $(\eta)$              & 5.8 & 16.9 \\
Primary dimensionless spin $(\chi_{1z})$          & 5.3 & 14.6 \\
Secondary dimensionless spin $(\chi_{2z})$          & 1.0 & 1.3 \\
Effective spin $(\chi_{\rm eff})$     & 4.9 & 13.4 \\
\hline
\hline
\end{tabular}
\caption{Improvement factor (as defined in Eq.~\ref{eq:improvement_factor}) relative to the quasi-circular case. Values greater than one indicate tighter constraints for the eccentric recovery. \textit{Note:} Parameters reported in the detector frame will have a 'det' subscript unless mentioned otherwise.}
\label{tab:intrinsic_improvement}
\end{table}

When accounting for eccentricity but neglecting finite-size effects, a compact binary is represented using 10 intrinsic parameters. 
These parameters are inherent to the binary system and do not depend on the observer. 
There are the two masses, the primary mass $m_{1}$ and secondary mass $m_{2}$, where $m_{1} \geq m_{2}$. 
The spin of each of the compact objects is encoded in a dimensionless spin vector $\vec \chi_i$, where $i=1,2$ denotes the object.
For this work, we limit our injections to systems with spins aligned with the orbital angular momentum, and we label the corresponding spin components $\chi_{i,z}$.
This removes two spin components per compact object, reducing the intrinsic parameter count by four.
Finally we have the two additional parameters to describe the elliptical orbit of the binary, the eccentricity $e$ and relativistic anomaly $\zeta$, both defined at a reference frequency.
We neglect finite-size effects such as the tidal deformability of the NS in this study.

In our analysis we find that the intrinsic parameters are increasingly better constrained as the injected eccentricity increases, although the improvements vary by parameters. 
Figure~\ref{fig:1D_intrinsic} shows the recovered marginalized 1D posteriors for several intrinsic parameters across the three injected eccentricities, illustrating the sharp tightening of the measurements with eccentricity, especially of the mass ratio, spin and eccentricity.
To quantify these eccentricity induced effects, we define the improvement factor for each measured parameter as
\begin{equation}
    \label{eq:improvement_factor}
    \mathcal{I} = \frac{\sigma_{\mathrm{circ}}}{\sigma_{\mathrm{ecc}}} \,.
\end{equation}
where $\sigma_{\mathrm{circ}}$ is the standard deviation of the marginalized posterior for the $e=0$ injection, and $\sigma_{\mathrm{ecc}}$ is the corresponding standard deviation for the eccentric injection.
We report $\mathcal I$ for various parameters for our runs in Tab.~\ref{tab:intrinsic_improvement}.
The precision of eccentricity measurement improves by a factor of $\mathcal{I}\sim 18$ for $\enineteen=0.25$ when compared to the quasi-circular case, although we note that the latter measurement is influenced by the prior boundary. This is what drives the massive improvement in the measurability of eccentricity. The corresponding improvement in eccentricity measurement between $\enineteen=0.25$ and $\enineteen=0.1$ is a factor of $\sim 6$.
In addition, the $\zeta$ posteriors are also increasingly informed by the data at higher eccentricity values, providing more information about the orbital configuration with increasing eccentricity. 

Our key result is that the mass and the spin parameters show significant precision improvements with eccentricity for the NSBH-like systems we study here. 
For the mass parameters, improvement is primarily observed in the detector-frame chirp mass $\mathcal{M}_{\mathrm{det}}$ up to a factor of $\mathcal{I}\sim 3$ and in mass ratio $q$ up to a factor of $\mathcal{I}\sim 20$ (see Fig.~\ref{fig:1D_intrinsic}). 
The improved measurement of $q$ in turn leads to improvement in the source-frame component masses. 
Interestingly, the secondary mass precision has a greater improvement than the primary mass. 
The only mass parameter with no improvement is the source-frame chirp mass $\mathcal{M}$. 
This can be attributed to the uncertainty in $\mathcal{M}$ being dominated by the precision of the redshift measurements, because the precision in $\mathcal{M}_\mathrm{det}$ is significantly smaller than $z$ ($\sigma_{\mathcal{M}_\mathrm{det}} \sim 10^{-3}$, $\sigma_z \sim 10^{-2}$). 
That combined with the lack of improvement found in redshift (see Sec.~\ref{sec:Extrinsic}) with increasing eccentricity drives the result. 
In addition to the mass parameters, we report improvement in the precision of $\chi_\text{eff}$ up to a factor of $\mathcal{I}\sim13$ leading to a similar order of improvement in $\chi_{1,z}$ up to a factor of $\mathcal{I}\sim15$.
Notably, for our $\rho_\mathrm{opt} = 20$ injections, we constrain $\chi_\mathrm{eff}<0$ for eccentricities $0.1$ and $0.25$ although the injected effective spin is small in magnitude, $\chi_\mathrm{eff} = -0.065$.

\begin{figure*}[htbp!]
    \includegraphics[width=0.7\textwidth]{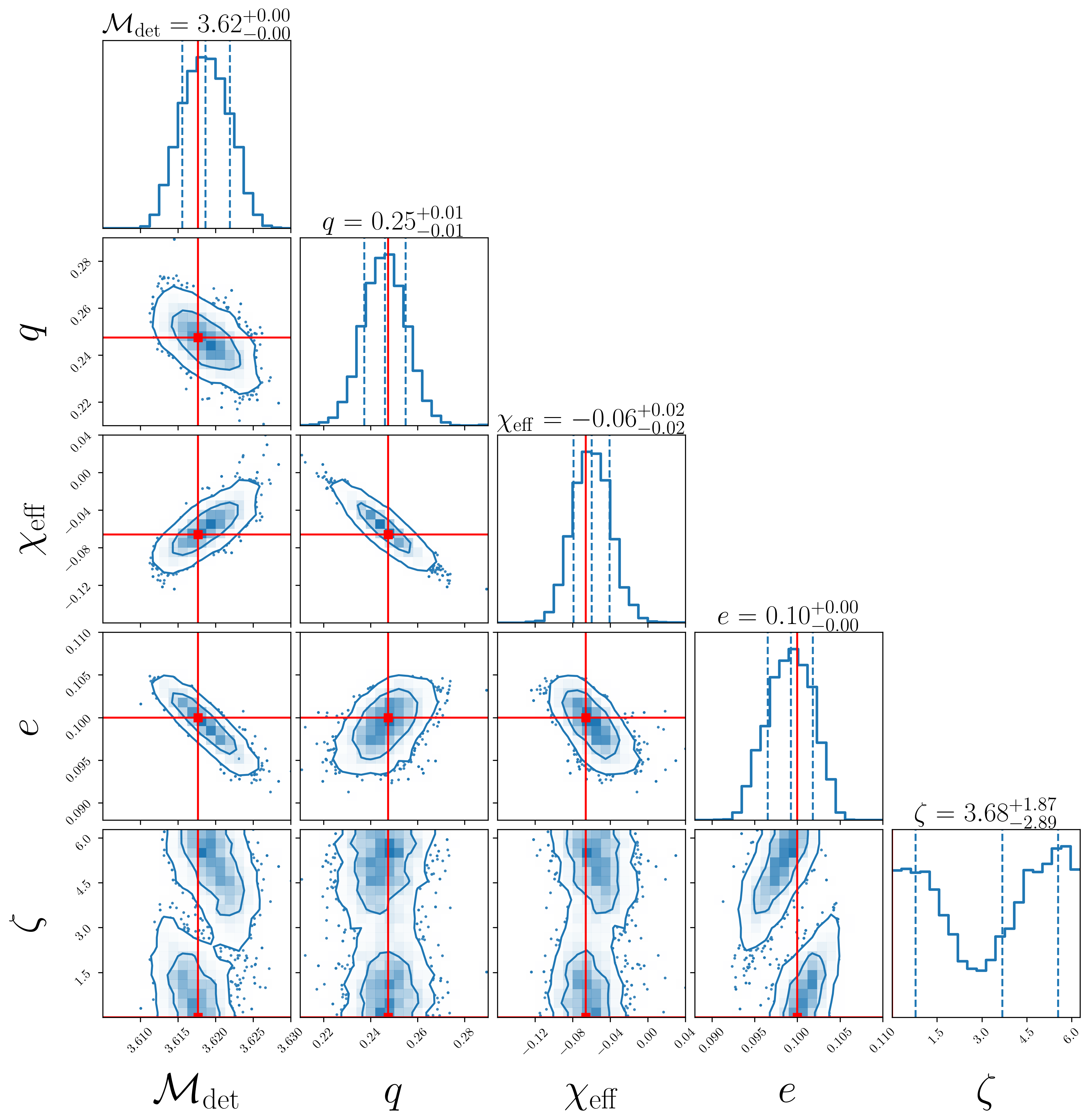}
    \caption{Corner plot for $\enineteen=0.1$ injection with 2D and 1D posteriors for the detector-frame chirp mass $\mathcal{M}_\mathrm{det}$, mass ratio $q$ (where $q=m_{2}/m_{1}$), effective spin parameter $\chi_\mathrm{eff}$, eccentricity $\enineteen$ and relativistic anomaly $\zeta_\mathrm{19}$. All posteriors are reported at the reference frequency, $f_\mathrm{ref}=19$ Hz. This plot clearly depicts the correlations between eccentricity and other intrinsic parameters.}
    \label{fig:Intrinsic1_Corner_e0.1}
\end{figure*}

To better understand the interplay of eccentricity and other intrinsic parameters, we focus on the $\enineteen=0.1$ injection. 
Fig.~\ref{fig:Intrinsic1_Corner_e0.1} shows a corner plot for the 2D and 1D marginalized posteriors for the case $\enineteen = 0.1$ for this case, exhibiting the clear correlations between eccentricity and all other intrinsic parameters.
The $e-\mathcal{M}$ and $e-\zeta$ posteriors show the strongest correlations, with the former due to the near-degeneracy between chirp mass and eccentricity in the lowest-order PN phasing~\cite{Favata:2021vhw}. 
We see milder correlation between eccentricity, and the mass ratio and effective spin. 
Finally, with significantly better constraints on mass and spin parameters, we see that increasing the magnitude of eccentricity does not break the the well known mass-spin degeneracy in the $q-\chi_\mathrm{eff}$ plane, although it tightens the constraints for these parameters. On the other hand, the 2D contours visibly shrink along the major axis in the $m_\mathrm{1} - m_\mathrm{2}$ plane. 
Plots for these and other results are given in Appx.~\ref{sec:appx_further_results}.

\subsection{Extrinsic parameter measurements}
\label{sec:Extrinsic}

\begin{figure*}[htbp!]
    \includegraphics[width=0.7\textwidth]{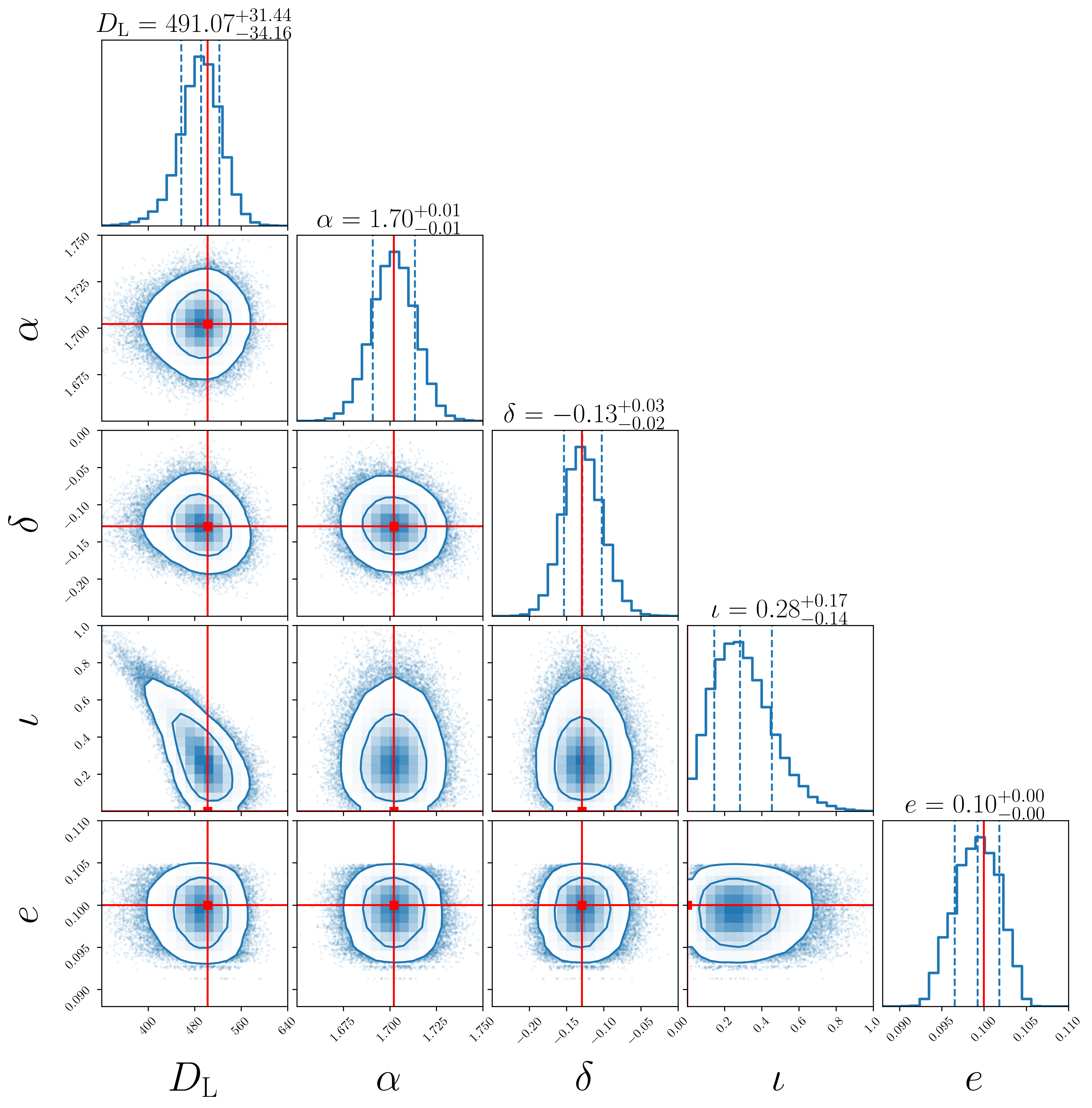}
    \caption{Corner plot for $\enineteen=0.1$ injection with 2D and 1D posteriors for the luminosity distance $D_\mathrm{L}$, right ascension $\alpha$, declination $\delta$, inclination angle $\iota$ and eccentricity $e$. All posteriors are reported at the reference frequency, $f_\mathrm{ref}=19$ Hz. The 2D contours display the lack of significant correlations between eccentricity and other extrinsic parameters.}
    \label{fig:Extrinsic_Corner_e0.1}
\end{figure*}

\begin{figure}[tb]
    \centering
    \includegraphics[width=0.48\textwidth]{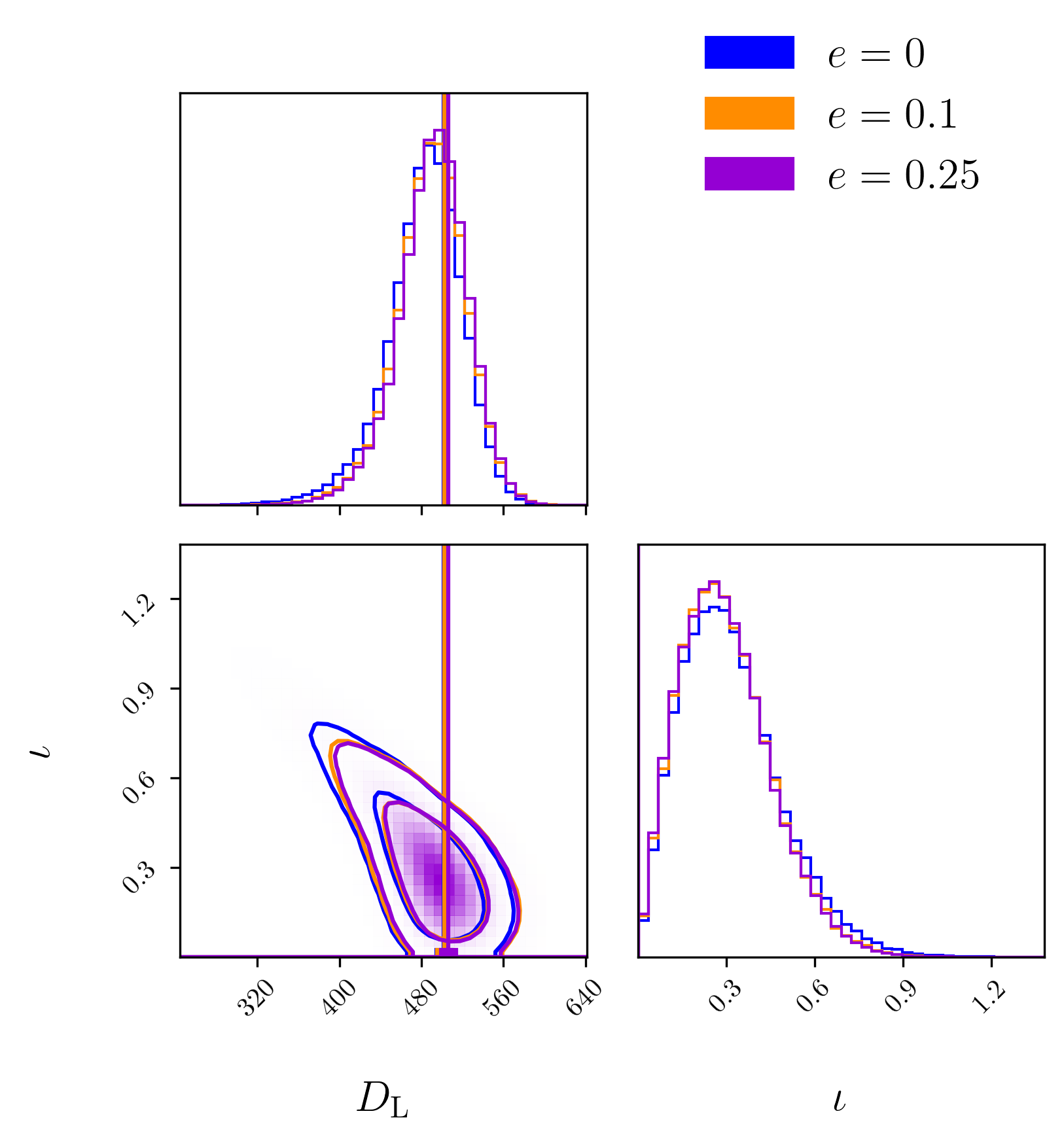}
    \caption{Corner plot for the luminosity distance $D_L$ and inclination $\iota$ for our face-on binaries, across injected eccentricities. All three signals are injected at varying distances such that the SNR for each injection is 20.0. However eccentricity causes only minute changes in the SNR and hence the distances are very close together.} The presence of nonzero eccentricity makes only a small change to the measurements.
    \label{fig:DL_iota}
\end{figure}

In addition to the intrinsic parameters, seven extrinsic parameters are used to describe a gravitational wave signal completely. 
These are luminosity distance $D_\mathrm{L}$, inclination angle $\iota$, sky localization characterized by right ascension $\alpha$ and declination $\delta$, polarization angle $\psi$, coalescence phase $\phi_\mathrm{c}$ and time $t_\mathrm{c}$. 
Fig.~\ref{fig:Extrinsic_Corner_e0.1} shows 2D and 1D marginalized posteriors for a subset of extrinsic parameters tied to localization and orientation of the GW source, in the case of $\enineteen= 0.1$.
As we use a three-detector network, the binary is relatively well-localized in sky position, while the luminosity distance remains correlated with the binary inclination through the overall waveform amplitude. 

For our NSBH-like binaries we find that eccentricity has a weak or negligible impact on the measurements in extrinsic parameter space, especially when compared to the effect on intrinsic parameters described in the previous section. 
For extrinsic parameters, the most noteworthy effect due to eccentricity is the modest tightening of the known $D_\mathrm{L}-\iota$ degeneracy. The 68\% credible interval of the highest-eccentricity case shows an improvement of up to 8.86\% (a factor of $\sim1.1$) in luminosity distance and 8.92\% (a factor of $\sim1.1$) in inclination, when compared to the circular case. 
These improvements are much smaller than the eccentricity-induced improvements in the intrinsic parameters described in Tab.~\ref{tab:intrinsic_improvement}. 
The same is illustrated in Fig.~\ref{fig:DL_iota}, showing the $D_\mathrm{L}-\iota$ posteriors across eccentricity values. 
A potential explanation for this effect is the additional orientation-dependence introduced by eccentricity, see e.g.~\cite{Yunes:2009yz}. 
In addition to this, we report precision improvement in measurement of $t_\mathrm{c}$ of up to 18\% (a factor of $\sim1.22$) for eccentricity values up to 0.25. 
The improvement factors (as described in Eq.~\eqref{eq:improvement_factor}) for all remaining extrinsic parameter recoveries are within $1^{+0.01}_{-0.01}$, categorizing them as being unaffected by the presence of eccentricity in our study of face-on systems.

\section{Follow-up Analysis}
\label{sec:Followup}

Up till here, we explore the effect of eccentricity on parameter measurements for face-on GW200105-like NSBH systems with higher modes included. In this next section, we assess the sensitivity of these trends to inclination and higher mode content in our injections.

\subsection{Inclination}
\label{sec:InclinationDependence}

\begin{figure*}[tb]
\centering
\includegraphics[width=0.48\textwidth]{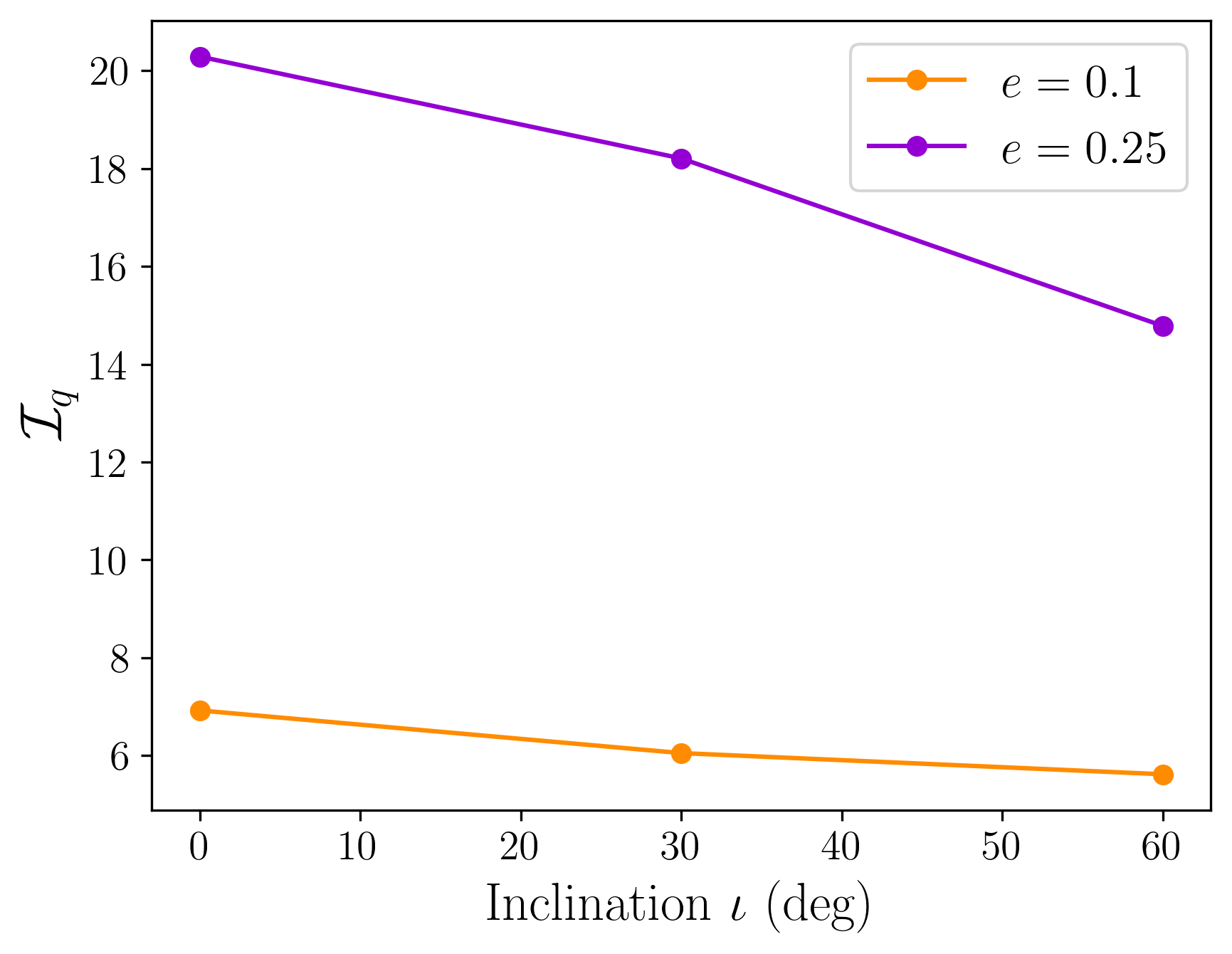}
\hfill
\includegraphics[width=0.48\textwidth]{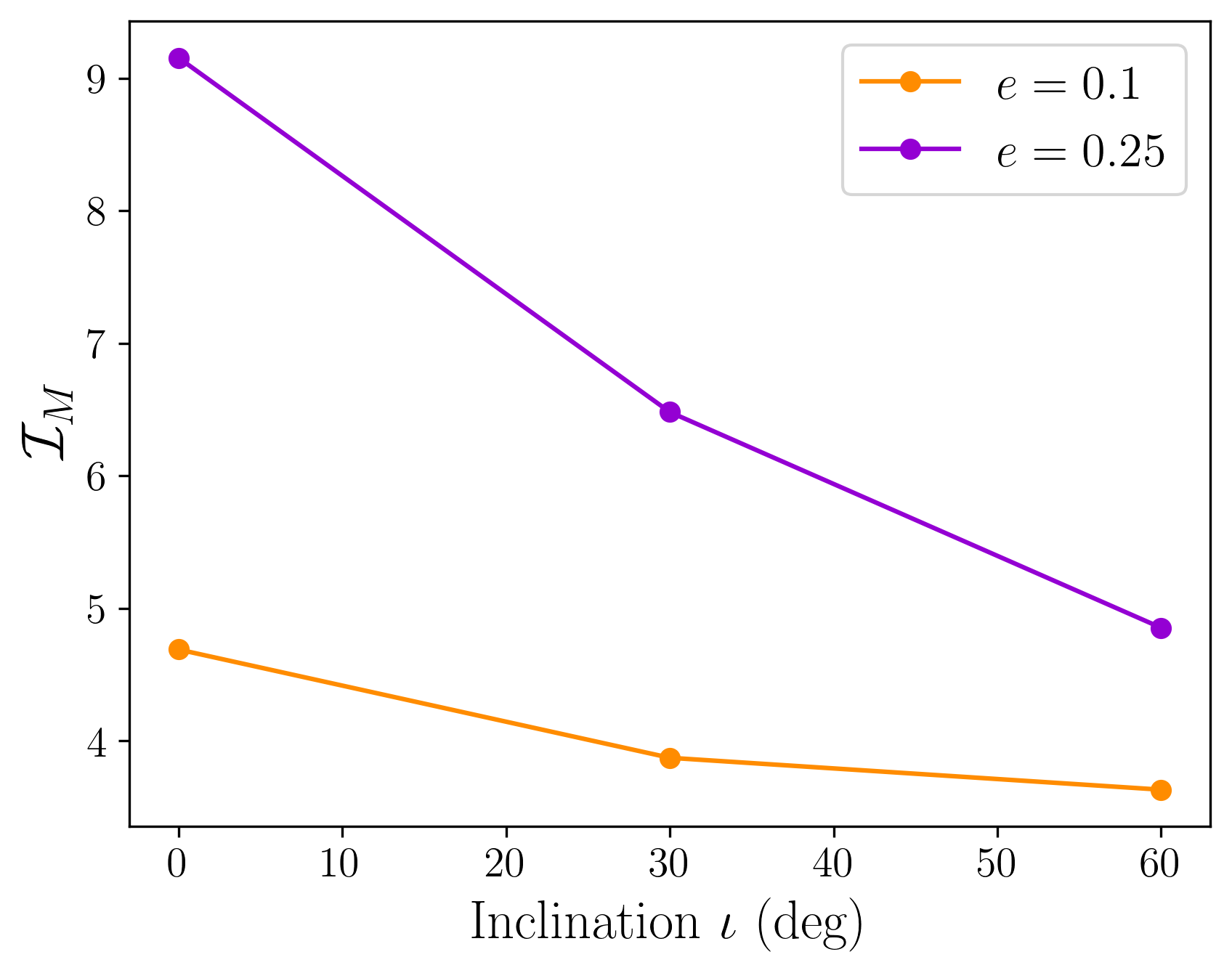}
\includegraphics[width=0.48\textwidth]{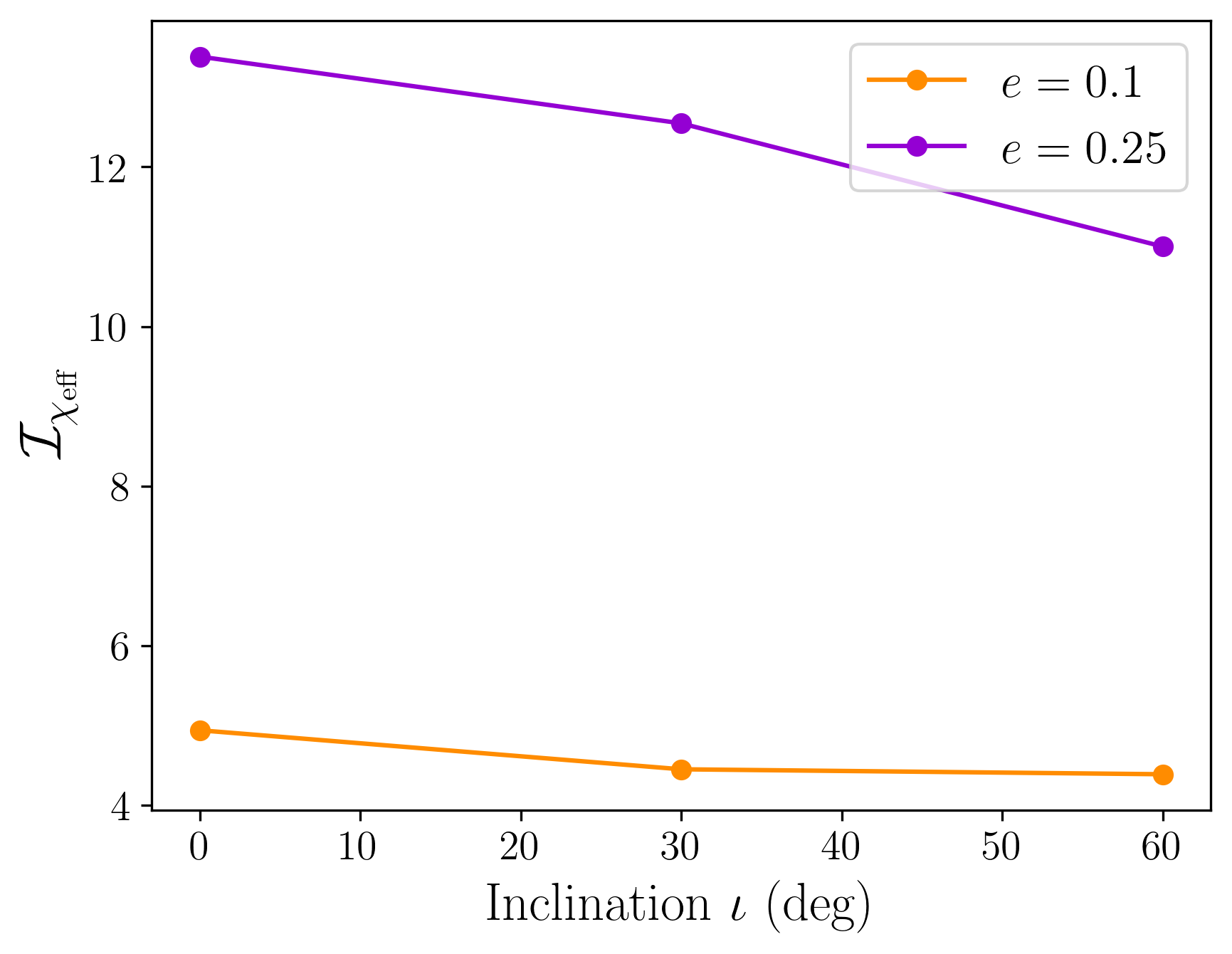}
\hfill
\includegraphics[width=0.48\textwidth]{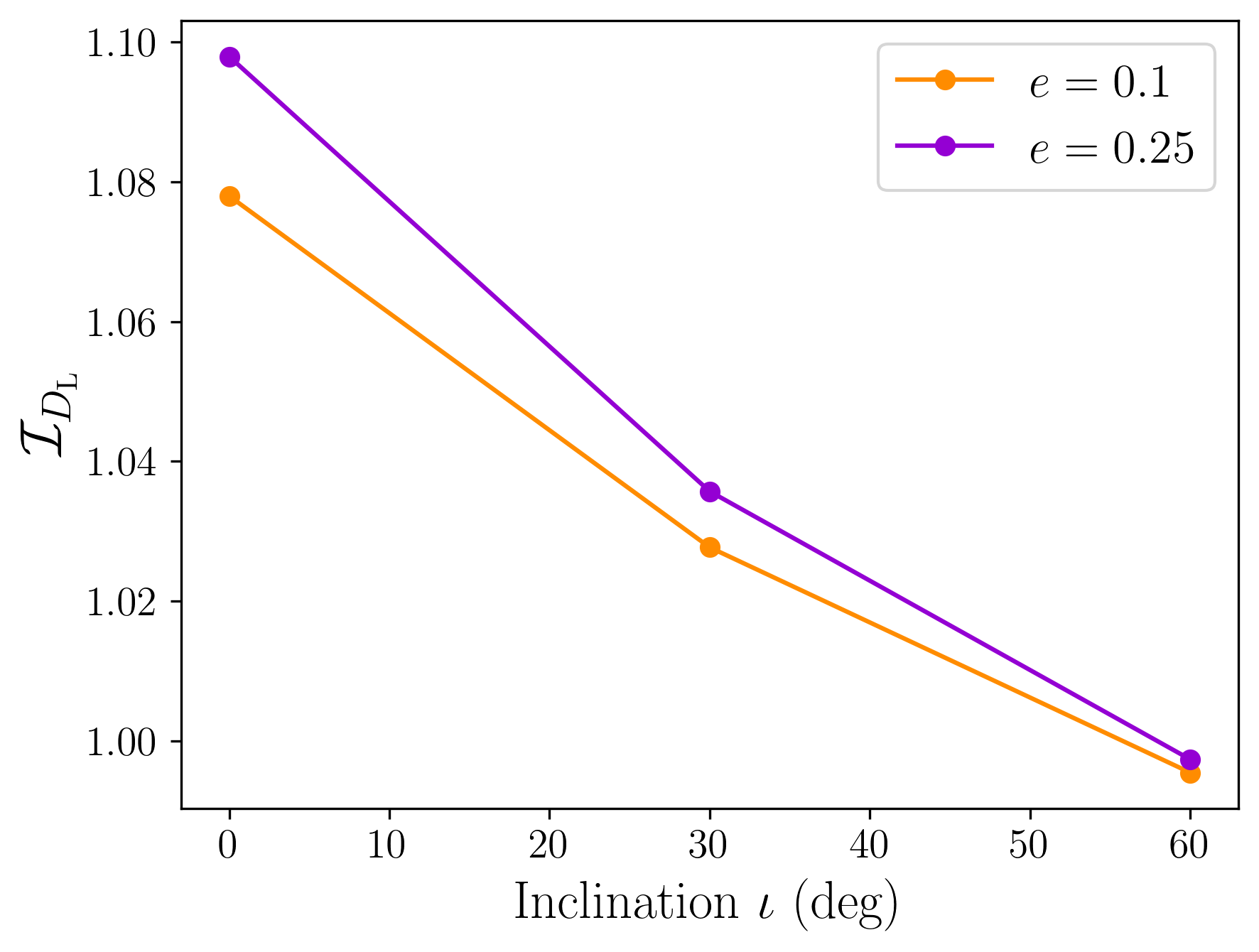}
\caption{Improvement factors, Eq.~\eqref{eq:improvement_factor} as a function of inclination for different measured parameters. These display the sensitivity of the eccentricity induced improvements to the various inclination angles. The plot showing improvement in luminosity distance (\textit{bottom-right}) may appear to be highly sensitive to inclination, but notice the scale variation on the y-axis.}
\label{fig:q_improv_v_incl}
\end{figure*}

We first carry out a suite of injection-recovery studies while varying the inclination angle $\iota$ of the system to the observer's line of sight.
The goal of this exercise is to better understand the inclination dependence of the eccentricity-induced improvement in measurement of parameters. 
Varying the inclination changes the relative amplitudes of emission from higher harmonics, and the leading eccentric harmonics can introduce new dependence on $\iota$ relative to the quasi-circular, leading quadrupole waveform~\cite{Yunes:2009yz}.
For this investigation, we vary inclination $\iota \in \{0, \pi/6, \pi/3\}$ along with varying the eccentricity as described in the previous sections, keeping all other parameters the same as mention in Tab.~\ref{table:fixed_param}. The results from this study are summarized in Fig.~\ref{fig:q_improv_v_incl}. 
We find that the eccentricity-induced improvements have a moderate to weak dependence on inclination.
The amount of improvement and the variation across inclinations depends on the intrinsic parameter considered.
However, the dependence on inclination value strengthens as the eccentricity increases across all parameters. 
For all parameters, the highest improvement is observed for the face-on configuration ($\iota=0$). 
This implies that the additional information provided by eccentricity is suppressed for binaries with inclination angles away from the face-on configuration.

\subsection{Higher-order modes}

\begin{figure}[h!]
    \centering
    \includegraphics[width=0.48\textwidth]{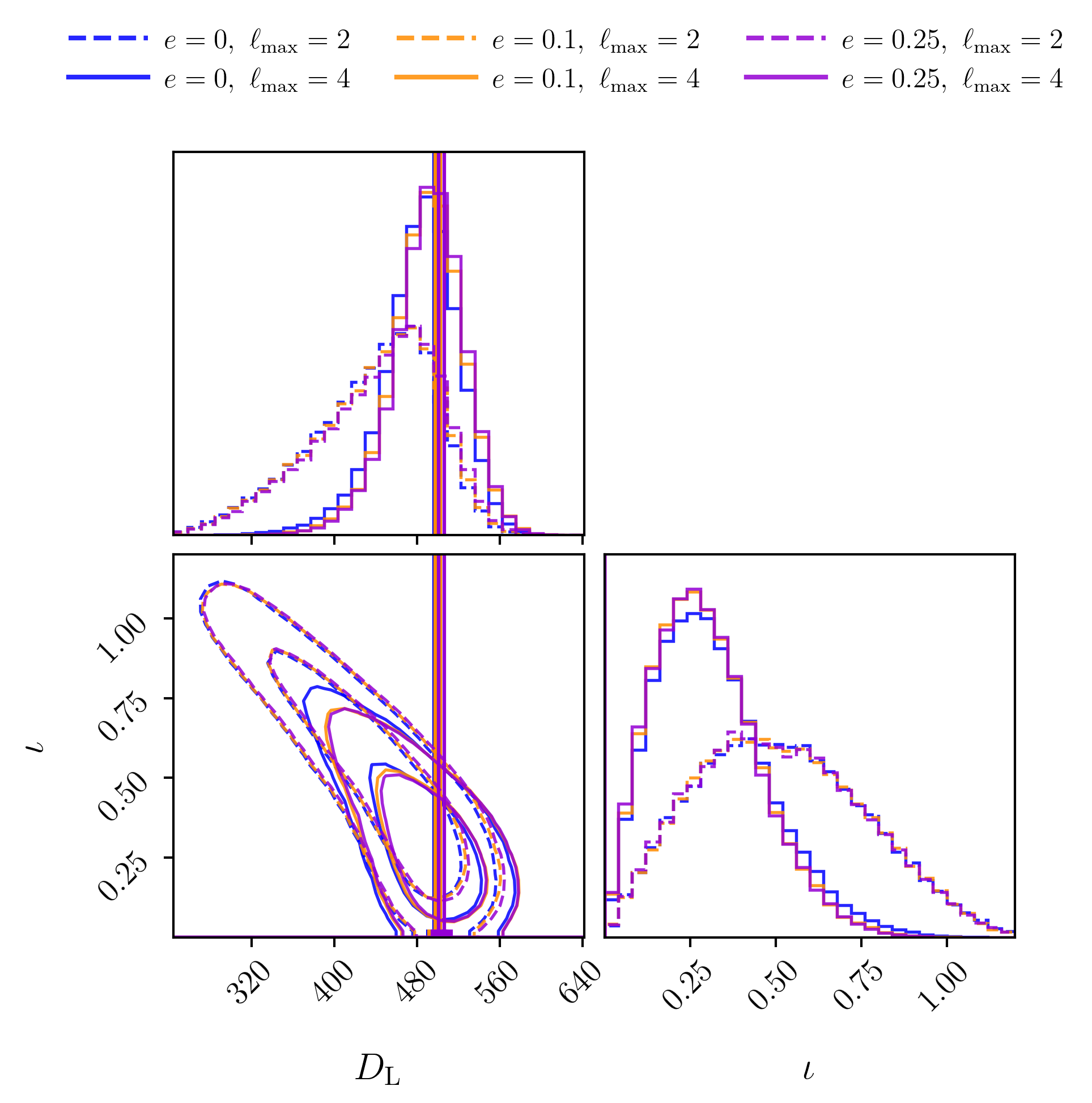}
    \caption{Similar to Fig.~\ref{fig:DL_iota}, corner plot for $D_L$ and $\iota$ while varying both eccentricity and higher mode content. 
    Solid lines include modes up to $\ell_\mathrm{max} = 4$, dashed lines include only quadrupolar emission, $\ell_\mathrm{max}=2$.
    While eccentricity makes very little difference in the measurement of these extrinsic parameters, higher modes enable better measurement along the major axis of the uncertainty contours as expected, resulting in a better measurement of both parameters.}
    \label{fig:HM_DL_iota}
\end{figure}
\label{sec:HM}
\begin{figure}
    \centering
    \includegraphics[width=0.48\textwidth]{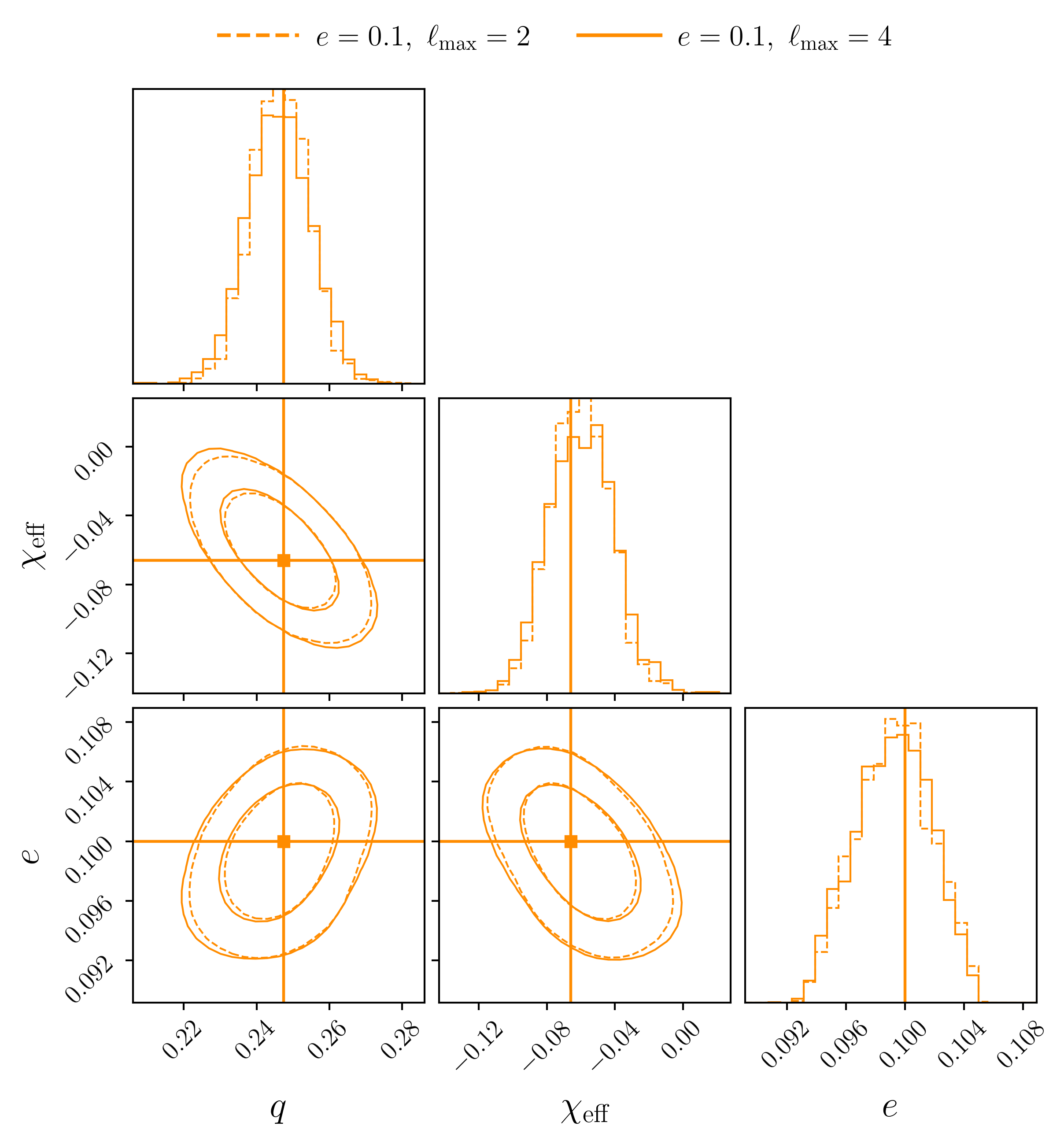}
    \caption{Corner plot for $q$, $\chi_\mathrm{eff}$ and $\enineteen$ while varying the higher mode content, for the $\enineteen=0.1$ face-on injection. 
    Solid lines include modes up to $\ell_\mathrm{max} = 4$, dashed lines include only quadrupolar emission, $\ell_\mathrm{max}=2$.
    Similar to the case of the extrinsic parameters, for this system the higher modes make only a small difference in the measurement of the intrinsic source parameters.}
    \label{fig:HM_intr}
\end{figure}

Higher-order modes from higher multipolar emission are known to break the distance-inclination degeneracy by introducing additional inclination dependence in the waveform that cannot be mimicked by a simple rescaling of the luminosity distance~\cite{LIGOScientific:2020stg, Kalaghatgi:2019log, OShaughnessy:2014shr, Usman:2018imj, Kumar:2018hml, London:2017bcn}.
They can also help in measuring the intrinsic parameters, as they contain information about the mass asymmetry of the system.
Our initial analysis included higher modes up to $\ell_\mathrm{max}=4$ and so to isolate the contribution of eccentricity from the information gained from the higher-order modes, we re-ran our analysis without the higher order modes present. 
{\it A priori} we expect the effects of higher-order modes to be small for our chosen injections, because although the NSBH-like systems have asymmetric masses, higher-order modes contribute more strongly at higher orbital velocities, towards the end of the long inspirals we study.
We find that for our injections, higher-order modes contribute only about $1-2\%$ of the total SNR.

In Fig.~\ref{fig:HM_DL_iota}, we compare the $D_\mathrm{L}-\iota$ corner plots for both the analyses (with and without higher modes). 
We find that including higher modes tightens our measurement of $D_L$ and $\iota$, as expected from the additional inclination dependence these modes introduce. 
However, we find that eccentricity has a small to negligible effect on the measurement of $D_L$ and $\iota$ in our NSBH-like systems, whether or not we include higher modes in our injection.
This indicates that for these component masses and values of eccentricity, the eccentric dynamics does not add significant information about the inclination.
This runs counter to what might be expected from the fact that eccentric harmonics can carry additional inclination dependence as compared to quadrupolar quasi-circular waveforms~\cite{Yunes:2009yz}.
However, this additional dependence enters the amplitude factors suppressed by the small value of eccentricity, and so does not add much information to the measurements. 

Meanwhile in Fig.~\ref{fig:HM_intr} we illustrate the effect of higher modes on the measurement of intrinsic parameters, for the case $\enineteen=0.1$. 
We find that the inclusion of higher modes makes negligible differences in our inferences, which is true across the three eccentricity values we study.
In this case, the higher modes are not breaking a strong degeneracy in the posteriors, and their low SNR contribution means they add little additional information.

\section{Discussion and Conclusions}
\label{sec:Discussion}

The observation of signatures of eccentricity in the NSBH binary that produced GW200105~\cite{morras2025orbitaleccentricityneutronstar} opens a new window onto compact binary systems that formed dynamically or have undergone recent dynamical interactions (see e.g.~\cite{Stegmann:2025clo}), resulting in measurable eccentricity near merger.
These systems lie in a surprising and thus far relatively under-explored part of binary parameter space: low mass, asymmetric systems, with potentially misaligned spins and orbital eccentricity.
Their characterization is particularly valuable because these parameters each retain information about the formation and subsequent dynamical evolution of the binary.

Our work quantifies the effects of this eccentricity on the parameter estimation of such GW200105-like NSBH systems. 
We find that the additional information provided by measurable eccentricity can yield much tighter constraints across the parameter space, especially the intrinsic parameters. 
This increasing precision in parameter recovery scales with the magnitude of eccentricity. 
The eccentricity-induced improvement of posteriors found in our results is consistent with the expectation that measurable eccentricity and associated parameters provide us with more information about the source system thus reducing uncertainty on all related parameters, but the degree of improvement for these systems is remarkable.
Through direct injection-recovery campaigns, we find that the intrinsic parameters of these systems can be measured several to more than ten times better than the corresponding quasi-circular case

This improvement can be understood through the additional structure that eccentricity imprints on the signal's time--frequency evolution.
Presence of orbital eccentricity gives rise to rapid variations in the orbital separation and velocity near periastron~\cite{Morras:2025nbp}.
This nonuniform orbital motion leads to the generation of eccentric harmonics, producing additional frequency components and sidebands that vanish in the quasi-circular limit~\cite{Peters:1963ux,Peters_1964,Patterson:2024vbo}.
In addition, the two-body system undergoes relativistic periastron precession, as predicted by general relativity~\cite{Damour:1985ca}.
These features help break the degeneracies between chirp mass, mass ratio, and effective spin — most notably for the source-frame masses, where the usual banana-shaped degeneracy gets progressively more compact and isotropic as eccentricity increases (see Appx.~\ref{sec:appx_further_results}).
This effect gets stronger with eccentricity, since the sidebands grow in amplitude and become more clearly distinguishable from what would otherwise look like small perturbations on a quasi-circular phase evolution.

Our results further indicate that much of this gain is already present in the dominant quadrupole contribution. The signal is strongly quadrupole dominated, with the $\ell_\mathrm{max}=2$ modes accounting for approximately $98.5\%$ of the total signal power in the configuration considered here. Eccentricity can therefore substantially sharpen the intrinsic-parameter posteriors even in the absence of significant higher mode contribution. In this sense, eccentricity and higher-order modes provide approximately complementary information: eccentric dynamics primarily enrich the phase and time--frequency evolution used to infer intrinsic parameters, whereas higher-order modes add angular and polarization structure that help constrain the viewing geometry.

This interpretation also explains why the improvements in the extrinsic parameters are substantially smaller. Luminosity distance and inclination remain strongly correlated because both affect the observed amplitude of the dominant radiation. Increasing eccentricity can strengthen subdominant waveform structure. However, for the systems and SNR considered here, it does not provide sufficient additional angular information to substantially break the distance--inclination degeneracy.
So while the intrinsic parameters tighten substantially with eccentricity, the distance--inclination correlation largely persists, yielding improvements of at most $\sim9\%$ for either parameter.

Higher-order modes can in principle break the distance--inclination degeneracy through their distinct angular dependence, but their impact here is limited by low signal content: modes beyond the $\ell_\mathrm{max}=2$ contribute only $\sim1.5\%$ of the total signal power. At a network SNR of 20, this corresponds to a higher-mode SNR of $\sqrt{0.015}\times20\simeq2.4$, suggesting that the angular information carried by these modes is only weakly measurable. This explains why including modes up to $\ell_\mathrm{max}=4$ produces only modest changes in the extrinsic parameters and no additional improvement in the intrinsic parameters.

This improved measurement in the mass and spin parameters of these systems means that as future observations uncover this population of eccentric NSBHs, each individual event is expected to carry a great deal of information.
For example, in our injections we assumed a small and negative effective spin, $\chi_\mathrm{eff} = -0.065$ (consistent with the maximum-likelihood point for GW200105~\cite{morras2025orbitaleccentricityneutronstar}), indicating that the spin component projected along the orbital angular momentum is, in aggregate, negative. Depending on the component-spin magnitudes, this may indicate spin--orbit misalignment or an anti-aligned component of the black-hole spin.
Such a small negative effective spin is challenging to measure, but at SNR 20 and with initial eccentricity $\enineteen = 0.25$, we confidently infer that $\chi_\mathrm{eff} < 0$ for our NSBH-like injection.
The recovery of the sign of such a small effective spin illustrates that eccentric waveform structure does more than improve the precision of already well-measured quantities: it can qualitatively alter the physical conclusions that can be drawn from an individual event. In a quasi-circular analysis, a posterior spanning both signs of $\chi_\mathrm{eff}$ may not distinguish aligned from antialigned spin contributions, whereas the eccentric signal considered here supports a definite negative projection.

Similarly, more precise measurements of individual binary properties can help distinguish formation scenarios with relatively fewer observations at a population level. The impact of our results on simulated population analysis and astrophysical inference is discussed further in Ref.~\cite{Salvarese:2026klm}.

These conclusions should nevertheless be interpreted within the scope of this study. Recent work done by Clarke et al.~\cite{Clarke:2026cuw} highlights that the interpretation and shape of eccentricity posteriors can be significantly sensitive to the waveform model, prior, and reference-frequency convention. 
Additionally, our zero-noise, fixed-SNR injections with a single waveform family isolate the information content of eccentricity but do not capture noise fluctuations or waveform systematics across various models. 
More generally, the quantitative improvements will depend on source masses, spin configuration, detector sensitivity, and reference eccentricity. 
The qualitative picture, however, remains robust: across all inclination angles, eccentricity drives significant improvements in intrinsic-parameter estimation, but gives only modest to no improvement for extrinsic parameters when higher-order-mode contributions are weak.

At the higher SNRs expected for nearby sources and next-generation detectors, both effects should strengthen. The eccentric time--frequency structure will further sharpen the intrinsic parameters, while the higher-mode content should become sufficiently measurable to better constrain the viewing geometry. Eccentric NSBH systems may therefore become especially information-rich sources.

\section*{ACKNOWLEDGMENTS}
We would like to thank Teagan Clarke for the LIGO Scientific Collaboration internal review, and Deirdre Shoemaker and Alberto Salverese for their helpful comments. This material is based upon work supported by NSF's LIGO Laboratory which is a major facility fully funded by the NSF. The authors are grateful for computational resources provided by LIGO Laboratory and supported by NSF Grants PHY-0757058 and PHY-0823459. 
ST was supported by NSF Grant PHY-2207780. 
AZ was supported by NSF Grant PHY-2308833. HYC is supported by NSF Grant PHY-2308752 and Department of Energy Grant DE-SC0025296.
The work was done by members of the Weinberg Institute and has an identifier of UT-WI-38-2025‌. 
The preprint number relevant to this work is LIGO-P2600328.

\appendix

\section{Prior details}
\label{sec:Priors}

Our priors are described in Table~\ref{tab:Priors}.
The priors on the spin components along the orbital angular momentum $\chi_{i,z}$ are the priors inherited for the components when assuming that the spin vectors are isotropic in direction and have a uniform prior on their magnitudes, $\chi_i \sim U[0,0.99]$~\cite{Ng:2018neg,Lange:2018pyp}. 

\begin{table}[t]

\label{tab:Priors}
\renewcommand{\arraystretch}{1.3}
\begin{tabular}{|c|c|c|}
\hline
\textbf{Parameter} & \textbf{Prior} & \textbf{Range} \\
\hline

$\mathcal{M}_\mathrm{det}$ & Uniform in component masses &
$3.6 - 3.64\,M_\odot$ \\
\hline

$\eta$ & Uniform in component masses &
$0.07 - 0.22$ \\
\hline

$\chi_{1,z}$ & Aligned-spin $z$prior &
$-0.99 - 0.99$ \\
\hline

$\chi_{2,z}$ & Aligned-spin $z$prior &
$-0.05 - 0.05$ \\
\hline

$\enineteen$ & Uniform &
$0 - 0.28$ \\
\hline

$\zeta$ & Uniform &
$0 - 2\pi$ rad \\
\hline

$d_L$ & Proportional to $d_L^2$ &
$1 - 750$ Mpc \\
\hline

$\iota$ & Proportional to $\sin\iota$ &
$0 - \pi$ rad \\
\hline

$\alpha$ & Uniform & $0 - 2\pi$ rad \\
\hline

$\delta$ & Proportional to $\cos\delta$ & $\frac{-\pi}{2} - \frac{\pi}{2}$ rad \\
\hline

$\psi$ & Uniform & $0 - \pi$ rad \\
\hline

$\phi_c$ & Uniform & $0 - 2\pi$ rad \\
\hline

$t_c$ & Uniform & $1000000000.0\pm\Delta$ \\
\hline

\end{tabular}
\caption{Prior ranges for parameters used in the analyses. $\Delta=0.075$ s}
\end{table}

\section{Further results and data tables}
\label{sec:appx_further_results}

\begin{figure*}[tb]
    \centering
    \includegraphics[width=0.48\textwidth]{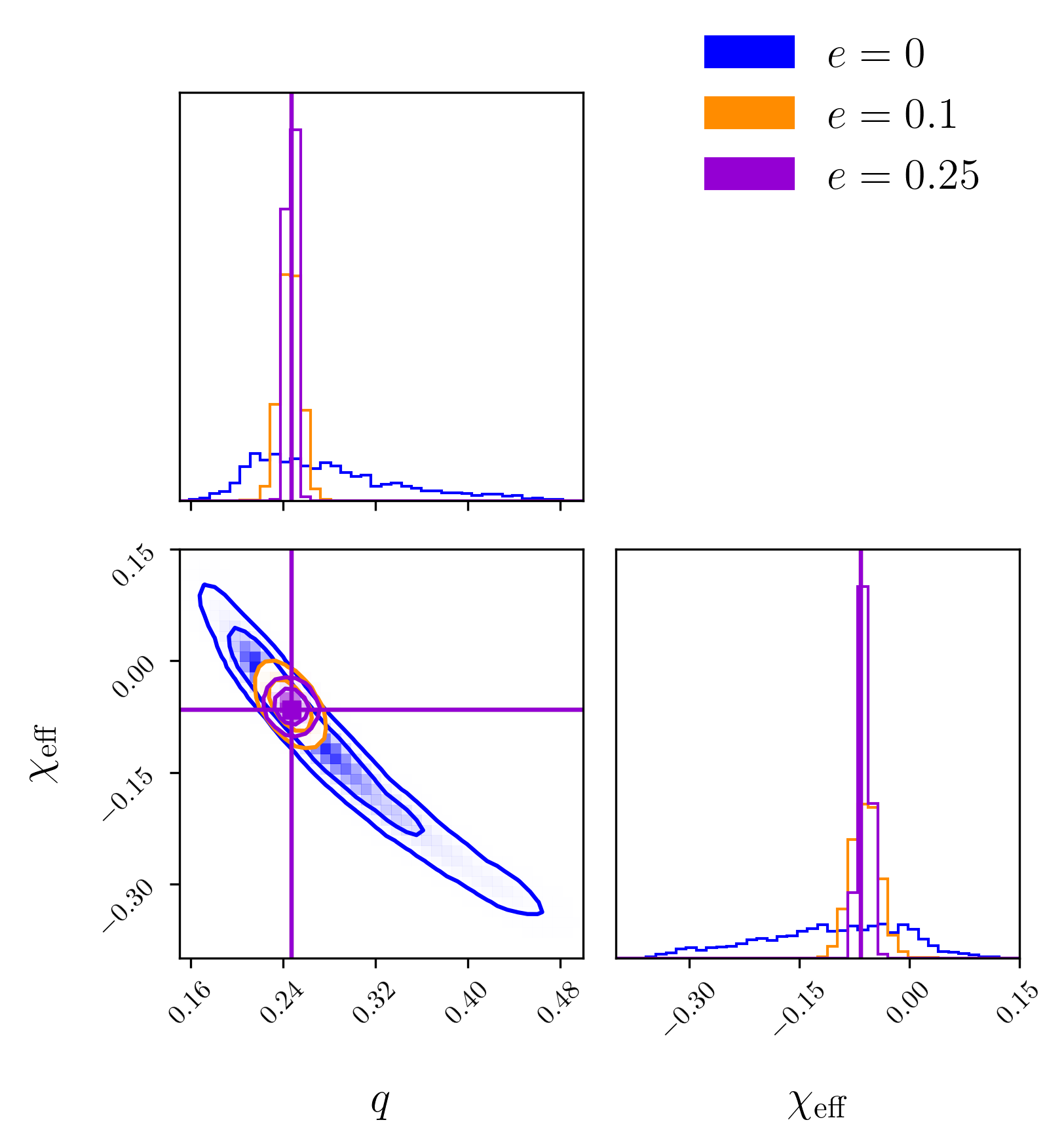}
    \hfill
    \includegraphics[width=0.48\textwidth]{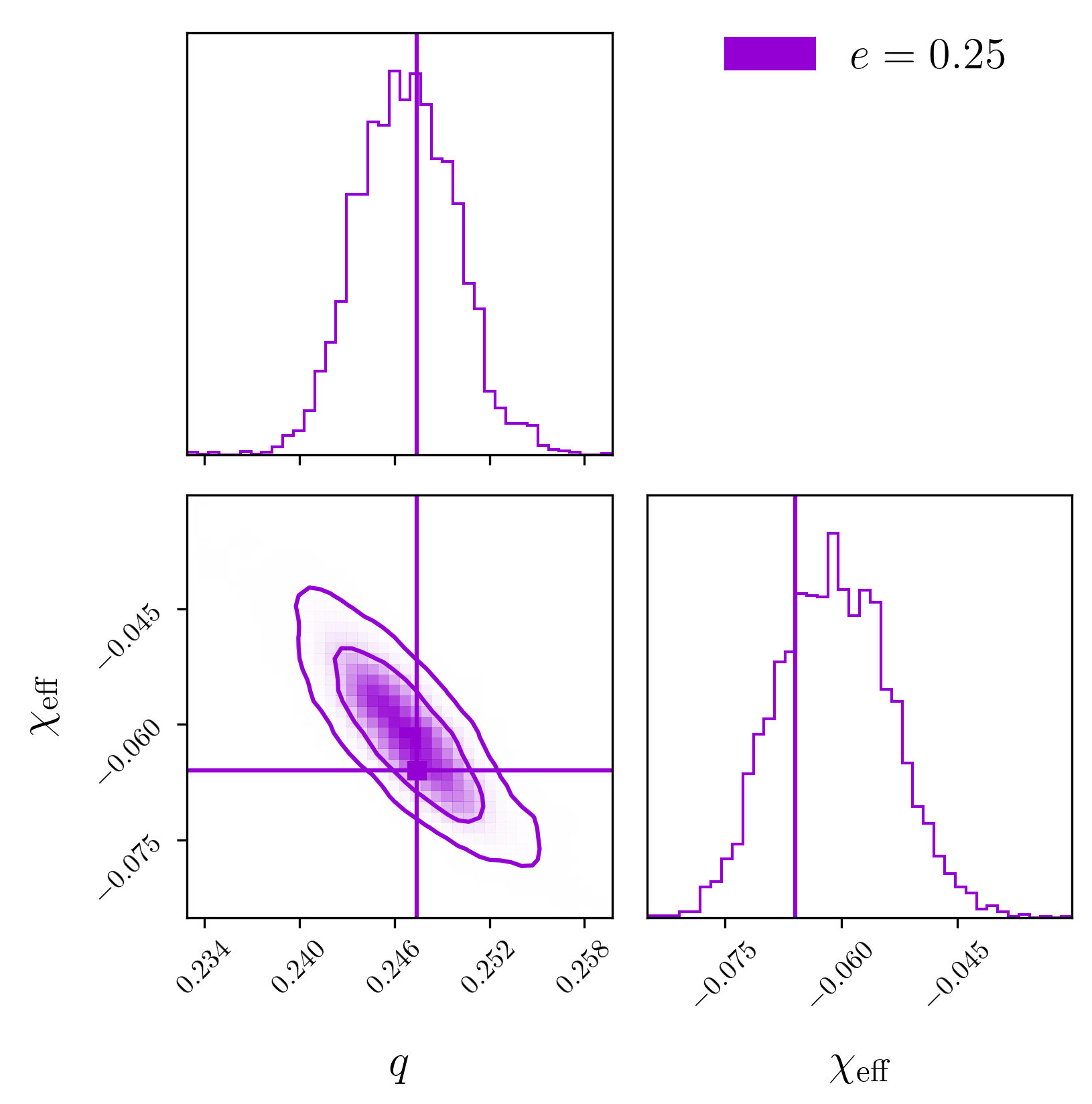}
    \caption{Corner plot showing the $q-\chi_\mathrm{eff}$ degeneracy across all three eccentricities (left) and specifically for $\enineteen=0.25$ (right). In the plot on the left, the degeneracy appears to have been broken by eccentricity; But when zoomed-in (right), the degeneracy still exists although in a much tighter region of the parameter space due to the decreasing uncertainty on $q$ and $\chi_\mathrm{eff}$ parameters in case of higher injected eccentricity.}
    \label{fig:chieff_q_degeneracy}
\end{figure*}

\begin{figure*}[tb]
    \centering
    \includegraphics[width=0.48\textwidth]{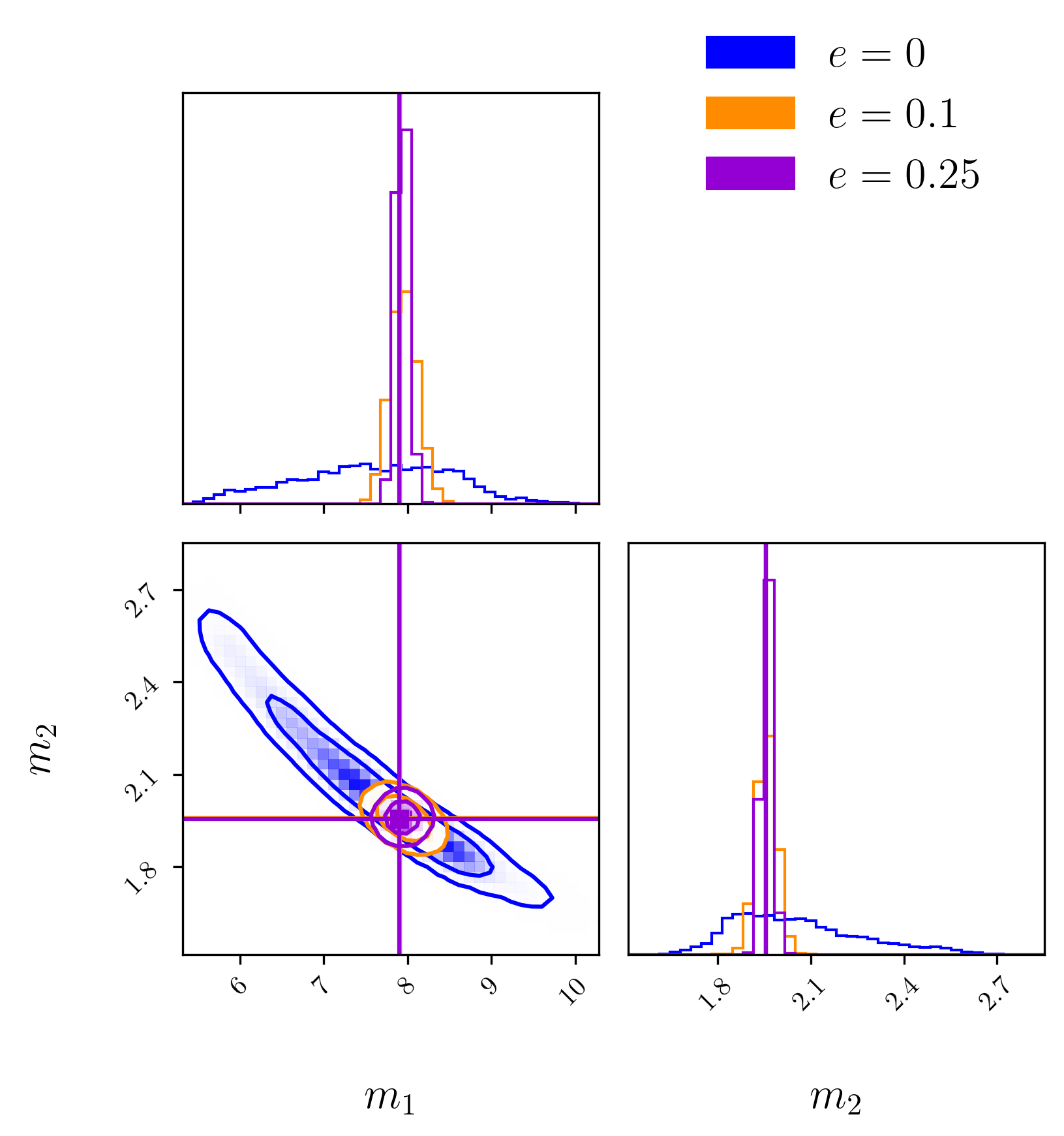}
    \includegraphics[width=0.48\textwidth]{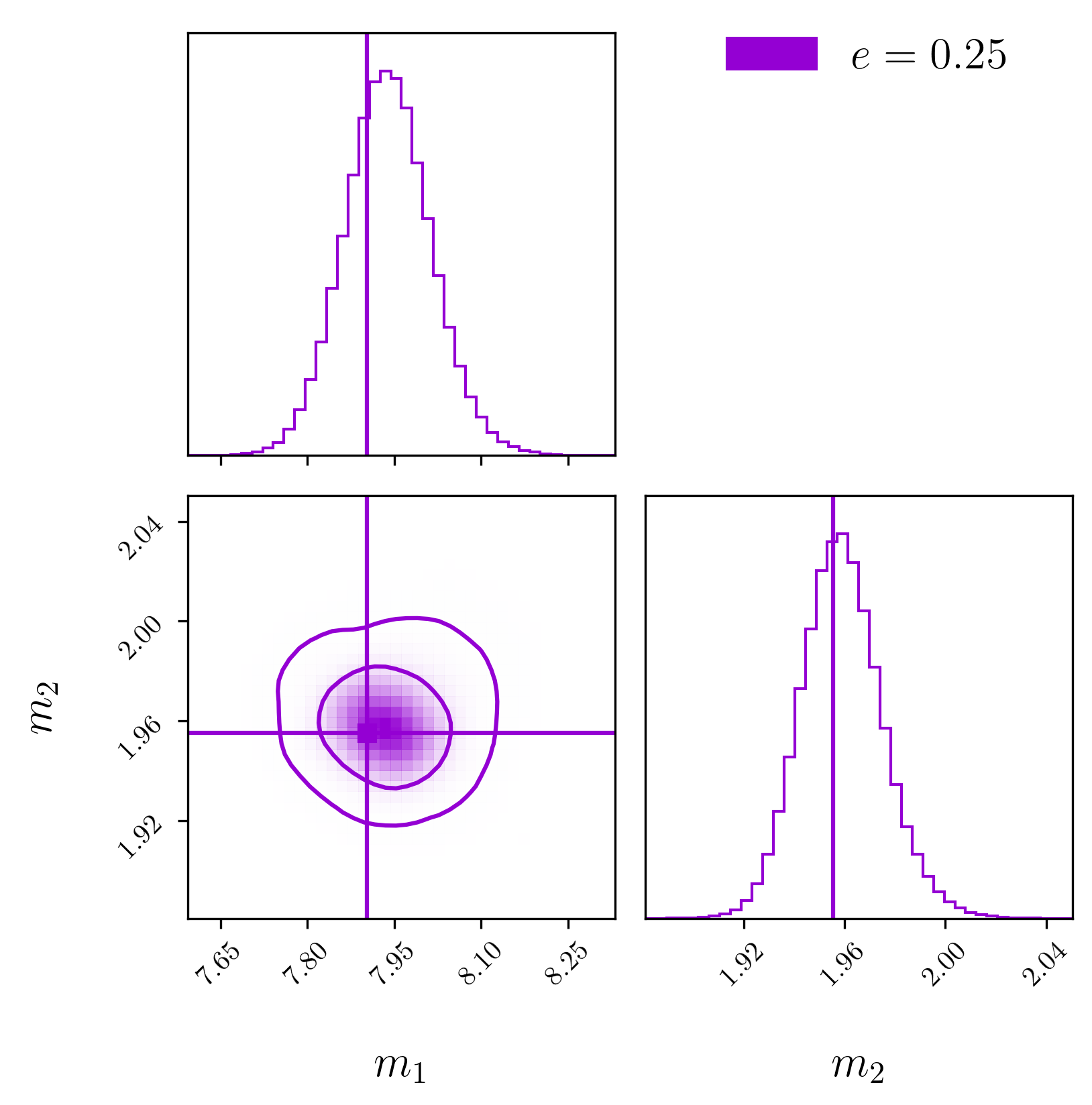}
    \caption{Measured source-frame component masses for the face-on injections discussed in Sec.~\ref{fig:1D_intrinsic}. Corner plot across all three eccentricities (left) showing greatly improved measurements with increasing eccentricity, and zooming in on the $\enineteen=0.25$ injection (right). Here we see visible compression in the degeneracy along the major axis of the banana-shaped correlation contour in case of higher injected eccentricity.
    } 
    \label{fig:m1_m2_degeneracy}
\end{figure*}

\begin{figure*}[tb]
    \includegraphics[width=0.48\textwidth]{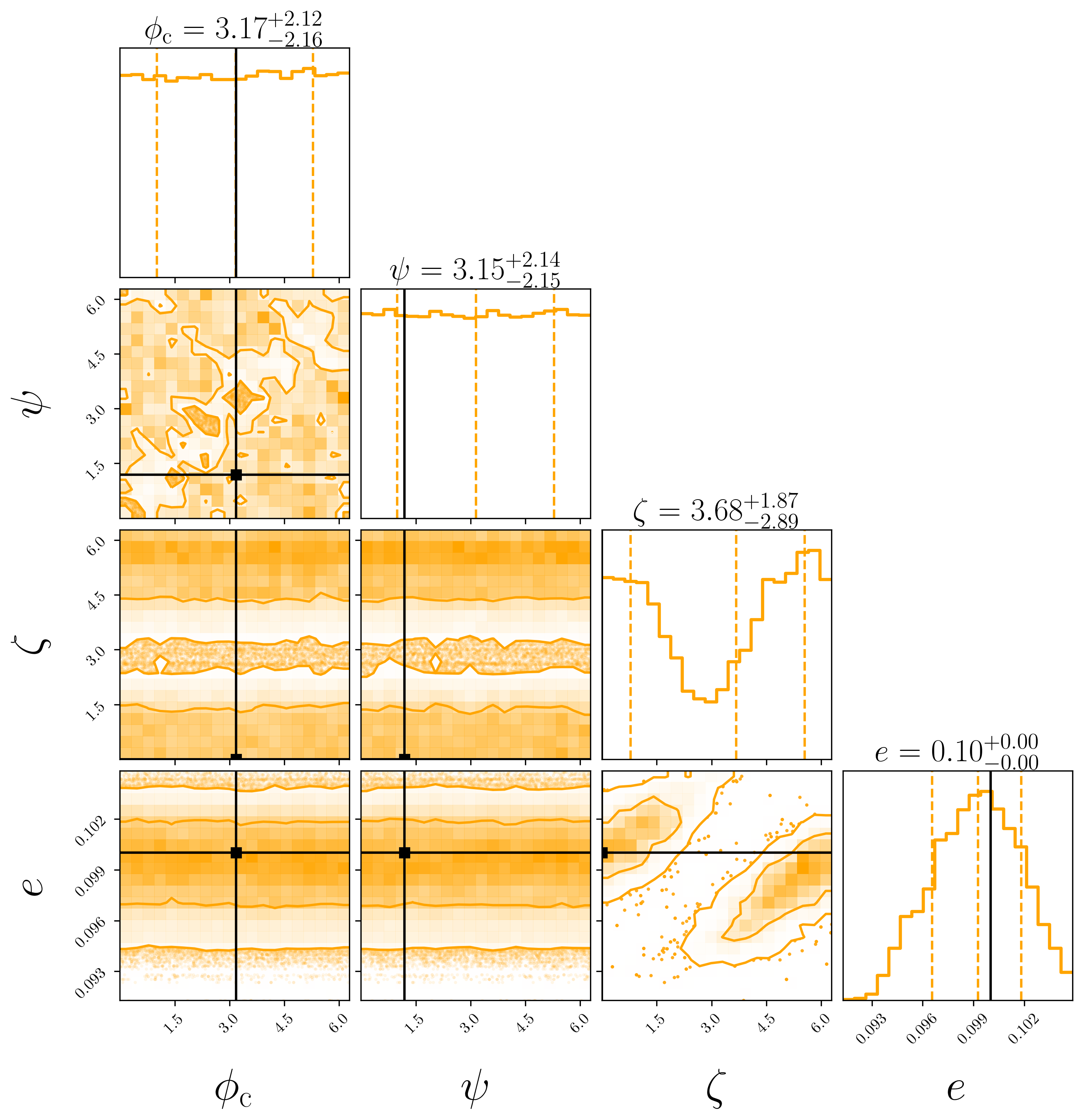}
    \includegraphics[width=0.48\textwidth]{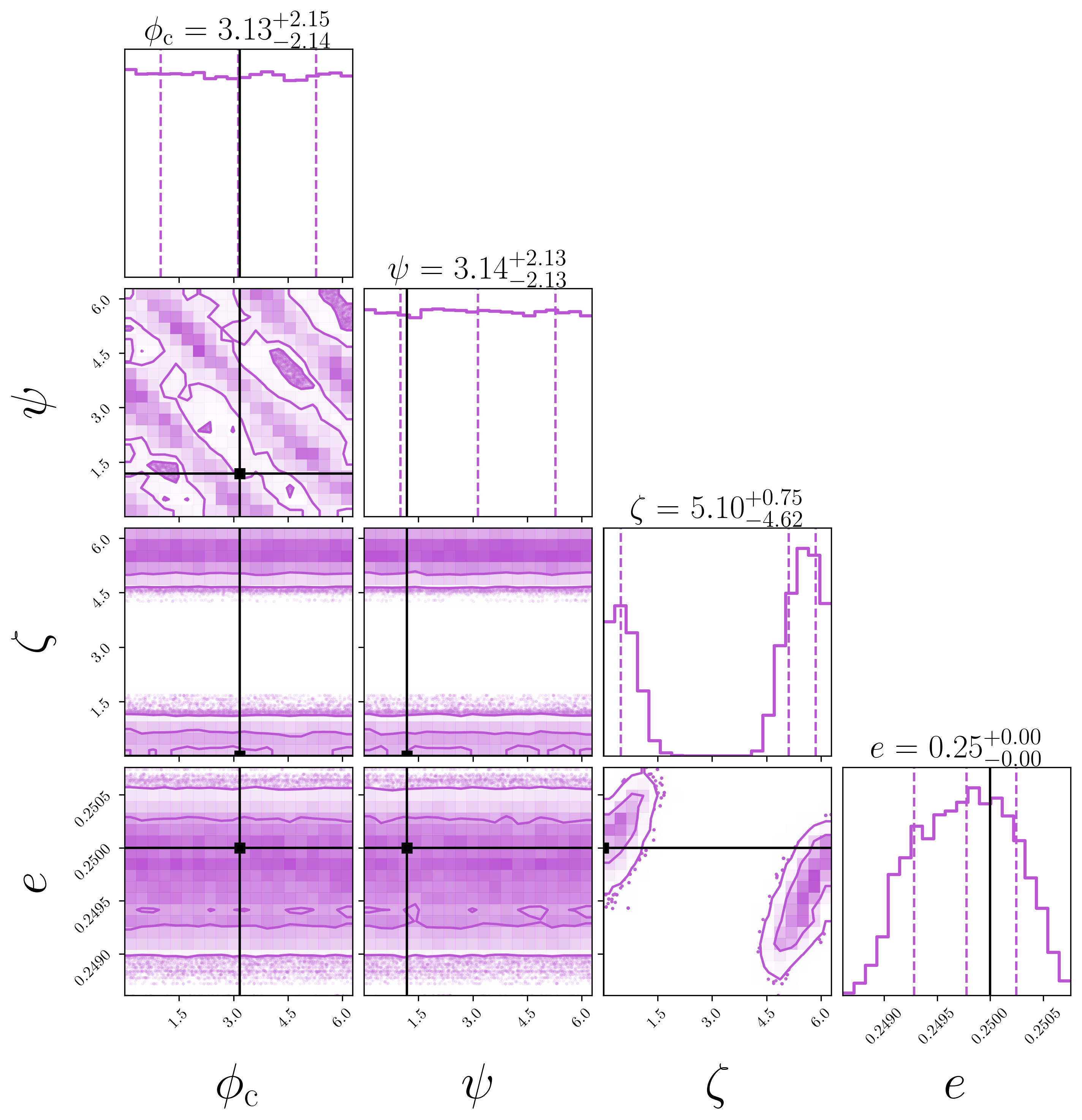}
    \caption{Corner plot with 2D and 1D posteriors for angular params for $\enineteen=0.1$ injection (left) and for $\enineteen=0.25$ injection (right). All posteriors are reported at the reference frequency, $f_\mathrm{ref}=19$ Hz. No unexpected correlations are observed here.}
    \label{fig:Angular_Corner_e0.1}
\end{figure*}

Here we give additional results and corner plots for our PE study.
Focusing first on the face-on injection suite, we compare the known $q-\chi_\mathrm{eff}$ degeneracy in Fig.~\ref{fig:chieff_q_degeneracy}. 
We see from the corner plot that this improvement is driven by the tighter measurement of the mass ratio $q$ and the primary aligned spin $\chi_\mathrm{1,z}$. 
While the tight measurements visually appear to lift the usual degeneracy between the component masses, we see that it persists even for $\enineteen=0.25$ (right panel of Fig.~\ref{fig:chieff_q_degeneracy}).
We also show a similar comparison for (source-frame) component mass measurements in Fig.~\ref{fig:m1_m2_degeneracy}.
The masses $m_\mathrm{1}$ and $m_\mathrm{2}$ see improvements $\mathcal I \sim 12$ for the highest eccentricity.
Here we observe a noticeable improvement in the degeneracy as the 2D contour for the highest eccentricity case shows a circularization of the usual oblong, banana shaped correlation contour(right panel of Fig.~\ref{fig:m1_m2_degeneracy}).
We further show illustrate the correlation between $e$, $\zeta$, and the extrinsic angular parameters $\phi_c$ and $\psi$ in Fig.~\ref{fig:Angular_Corner_e0.1}, for both non-zero eccentricity cases.

Lastly, Table~\ref{table:int_error} gives the standard deviation computed from the 1D marginalized posterior samples for each measured intrinsic parameter, across all our PE runs.
Table~\ref{table:ext_error} gives the same information for the extrinsic parameters.

\begin{table*}[b]
\caption{Standard deviation of posteriors for intrinsic parameters for all injections.}
\begin{ruledtabular}
\begin{tabular}{c|ccccccc}
\multicolumn{8}{c}{$\iota$ = 0} \\
\hline
$\enineteen$ & $M$ & $\mathcal{M}_\mathrm{det}$ & $q$ & $\chi_\mathrm{1,z}$ & $\chi_\mathrm{2,z}$ & $\chi_\mathrm{eff}$ & $\enineteen$\\
\hline
0.00 & 0.702771 & 0.004438 & 0.063502 & 0.134328 & 0.016034 & 0.097292 & 0.007740 \\
0.10 & 0.149840 & 0.002961 & 0.009171 & 0.025151 & 0.015401 & 0.019698 & 0.002503 \\
0.25 & 0.076778 & 0.001619 & 0.003131 & 0.009176 & 0.012162 & 0.007270 & 0.000435 \\
\hline
\multicolumn{8}{c}{$\iota$ = $\pi/6$} \\
\hline
$\enineteen$ & $M$ & $\mathcal{M}_\mathrm{det}$ & $q$ & $\chi_\mathrm{1,z}$ & $\chi_\mathrm{2,z}$ & $\chi_\mathrm{eff}$ & $\enineteen$\\
\hline
0.00 & 0.690401 & 0.004219 & 0.058353 & 0.125355 & 0.016549 & 0.092104 & 0.007328 \\
0.10 & 0.178245 & 0.003000 & 0.009641 & 0.026057 & 0.015360 & 0.020696 & 0.002545 \\
0.25 & 0.106529 & 0.001556 & 0.003206 & 0.009350 & 0.012166 & 0.007341 & 0.000422 \\
\hline
\multicolumn{8}{c}{\textbf{$\iota$ = $\pi/3$}} \\
\hline
$\enineteen$ & $M$ & $\mathcal{M}_\mathrm{det}$ & $q$ & $\chi_\mathrm{1,z}$ & $\chi_\mathrm{2,z}$ & $\chi_\mathrm{eff}$ & $\enineteen$\\
\hline
0.00 & 0.651444 & 0.003804 & 0.047429 & 0.106239 & 0.015936 & 0.080472 & 0.006270 \\
0.10 & 0.179319 & 0.002967 & 0.008439 & 0.023095 & 0.015320 & 0.018339 & 0.002505 \\
0.25 & 0.134202 & 0.001569 & 0.003208 & 0.009203 & 0.012209 & 0.007314 & 0.000428 \\
\end{tabular}
\end{ruledtabular}
\label{table:int_error}
\end{table*}

\begin{table*}[b]
\caption{Standard deviation of posteriors for extrinsic parameters for all injections.}
\begin{ruledtabular}
\begin{tabular}{c|cccccc}
\multicolumn{7}{c}{$\iota$ = 0} \\
\hline
$\enineteen$ & $\iota$ & $D_\mathrm{L}$ & $\alpha$ & $\delta$ & $\phi_{c}$ & $t_\mathrm{c}$ \\
\hline
0.00 & 0.168994 & 37.973485 & 0.011338 & 0.025670 & 1.809873 & 0.000406 \\
0.10 & 0.154727 & 35.226694 & 0.011396 & 0.025635 & 1.816774 & 0.000347 \\
0.25 & 0.154018 & 34.586160 & 0.011195 & 0.025285 & 1.817962 & 0.000332 \\
\hline
\multicolumn{7}{c}{$\iota$ = $\pi/6$} \\
\hline
$\enineteen$ & $\iota$ & $D_\mathrm{L}$ & $\alpha$ & $\delta$ & $\phi_{c}$ & $t_\mathrm{c}$ \\
\hline
0.00 & 0.213246 & 54.244175 & 0.011319 & 0.025751 & 1.412318 & 0.000401 \\
0.10 & 0.206855 & 52.783301 & 0.012460 & 0.023222 & 1.755149 & 0.000325 \\
0.25 & 0.204815 & 52.374849 & 0.011193 & 0.025114 & 1.141837 & 0.000325 \\
\hline
\multicolumn{7}{c}{$\iota$ = $\pi/3$} \\
\hline
$\enineteen$ & $\iota$ & $D_\mathrm{L}$ & $\alpha$ & $\delta$ & $\phi_{c}$ & $t_\mathrm{c}$ \\
\hline
0.00 & 0.205446 & 66.413936 & 0.011114 & 0.025386 & 1.203230 & 0.000368 \\
0.10 & 0.205947 & 66.718574 & 0.011059 & 0.025238 & 1.710256 & 0.000324 \\
0.25 & 0.203965 & 66.591831 & 0.010925 & 0.024616 & 1.003826 & 0.000296 \\
\end{tabular}
\end{ruledtabular}
\label{table:ext_error}
\end{table*}

\bibliography{references.bib}

\end{document}